\documentclass[a4paper, 11pt]{article}
\usepackage[authoryear,super]{natbib}
\usepackage[english]{babel}
\usepackage{pgf}
\usepackage{graphicx}
\usepackage{url,booktabs,multirow}
\usepackage{hhline}
\usepackage{amsmath,amsfonts,amssymb,amsbsy,amsthm,bm}
\usepackage{nicefrac}
\usepackage{subfigure}
\usepackage{color}
\usepackage{times}
\usepackage{multicol}
\usepackage{setspace}
\usepackage[utf8]{inputenc}
\usepackage{float}
\usepackage{bbold} 
\usepackage{algorithm} 
\usepackage{algpseudocode}
\usepackage{bm}
\usepackage{bbm}
\usepackage{algorithm}

\usepackage{tikz} 
\usetikzlibrary{shapes.geometric, shapes.multipart, arrows} 

\tikzstyle{startstop} = [rectangle, rounded corners, minimum width=3cm, minimum height=0.75cm,text centered, draw=black, fill=white]
\tikzstyle{process} = [rectangle, minimum width=3cm, minimum height=0.75cm, text centered, draw=black, fill=white]
\tikzstyle{arrow} = [thick,->,>=stealth, rounded corners]

\onehalfspace
\usepackage[a4paper,left=2.5cm,right=2.5cm,top=2cm,bottom=2cm]{geometry}

\usepackage{authblk}

\def\ci{\perp\!\!\!\perp}

\author[*]{Paul Messer}
\author[*]{Timo Schmid}

	\definecolor{bleudefrance}{rgb}{0.19, 0.55, 0.91}

\affil[*]{\small{Institute of Statistics, Otto-Friedrich-Universität Bamberg}}
\date{}

\begin{document}
\title{Gradient Boosting for Hierarchical Data in Small Area Estimation}
\maketitle

\begin{abstract}
This paper introduces Mixed Effect Gradient Boosting (MEGB), which combines the strengths of Gradient Boosting with Mixed Effects models to address complex, hierarchical data structures often encountered in statistical analysis. The methodological foundations, including a review of the Mixed Effects model and the Extreme Gradient Boosting method, leading to the introduction of MEGB are shown in detail. It highlights how MEGB can derive area-level mean estimations from unit-level data and calculate Mean Squared Error (MSE) estimates using a nonparametric bootstrap approach. The paper evaluates MEGB’s performance through model-based and design-based simulation studies, comparing it against established estimators. The findings indicate that MEGB provides promising area mean estimations and may outperform existing small area estimators in various scenarios. The paper concludes with a discussion on future research directions, highlighting the possibility of extending MEGB's framework to accommodate different types of outcome variables or non-linear area level indicators.
\end{abstract}

\newpage
\section{Introduction}
Small area estimation (SAE) models can be split into two groups, \textit{unit-level} models and \textit{area-level} models. As can be deferred from the names, unit-level models use data at the individual (micro) level, whereas area-level models only use aggregated data.  An example of a unit-level model is the Battese-Harter-Fuller (BHF) \citep{battese}, whereas an example of an area-level model is the Fay-Herriot model.
 One advantage of using unit-level survey data is, that it provides more detailed information than area-level data, this goes along with a considerably larger sample size. 
Estimators such as the Empirical Best Predictor (EBP) \citep{molina2009small} are based on unit-level auxiliary data. Another unit-level estimator is the MERF \citep{MERF}.
The MERF integrates the power and flexibility of machine learning methods with random effects, which has the advantage that the method does not rely on many model assumptions and combines the advantages of Random Forests with the ability to control for cluster structures of survey data.

 \cite{Krennmair2022} extend the MERF to the context of SAE and encourage the use of further machine learning methods such as Support Vector Machines or Gradient Boosting. As Gradient Boosting is often expected to achieve better prediction results than Random Forests, Gradient Boosting combined with Mixed Effects will be the subject of this paper.  
So far, Gradient Boosting was only used in the context of small area estimation without random effects \citep{merfeld_newhouse_xgboost}. 
Therefore, this paper proposes the estimation model Mixed Effect Gradient Boosting, a small area model, which is capable of handling different levels of complexity and flexibility through non-linearity and a nonparametric approach. 

Section \ref{methods} introduces the used methods and concepts, where Mixed Effects and Gradient Boosting are briefly explained. This is followed by an introduction of the new proposed method Mixed Effect Gradient Boosting, which combines the previously mentioned estimation methods through an Expectation-Maximization (EM) algorithm. Next, the same section shows how to derive area-level mean estimations from unit-level estimations. Furthermore, Section \ref{uncertainty} includes how to calculate MSE estimates for the estimated area means by using a non-parametric bootstrap.


For unit-level models, such as the MERF, the EBP with box-cox transformation (EBP-BC) \citep{rojas2020data}, or the newly introduced MEGB, certain data requirements have to be met. Unit-level data from a survey (for instance at household level, which is the case in the design-based simulation study in Section \ref{db}) requires covariates as well as observations of the variable of interest. In contrast, for the unit-level population data only the covariates  $\boldsymbol{x}$ are needed.
In Western countries, micro-level population data often cannot be used in design-based simulations, therefore additional
data sources are needed and used.



Furthermore, the performance of the proposed method is compared to established estimators and evaluated via a model-based and a design-based simulation study in sections \ref{mb} and \ref{db} respectively. These sections also include an evaluation of the MEGB based on the MSE estimates, which were created using a non-parametric bootstrap. The MERF in particular is used as a benchmark to assess the performance of the MEGB. To achieve comparability between the MEGB and the MERF, the simulation studies follow those described in \cite{Krennmair2022}. Subsequently, the results are discussed in subsections \ref{mb_results} and \ref{db_results}. Possible suggestions and motivation for further research are proposed and closing remarks follow in Section \ref{conclusion}.



\section{Concepts and Methods}
\label{methods}

This chapter introduces the different concepts and methods used to define and understand the MEGB, which itself is defined in Section \ref{Mixed effects gradient boosting}.

\subsection{Review of Linear Mixed Effect Models}
\label{mixed_effects}
Linear mixed models (LMM) include random effects in addition to the fixed effects $f(\boldsymbol{X})$. Mixed models enable the modeling and estimation of area-specific effects, especially in the case of small sizes in subsamples from areas $d = 1, ... , D$. In this regard, the linear effect $f(\boldsymbol{X})$ is extended to $f(\boldsymbol{X}) + \boldsymbol{Z} \boldsymbol{\vartheta}$. Here $\boldsymbol{Z}$ is a subvector of the covariates and $\boldsymbol{\vartheta}$ is a vector of the area-specific random effects.
Models solely based on fixed effects are usually not entirely suitable for this purpose since a too large number of free parameters has to be estimated compared to the sample size. An advantage of mixed models is, that correlations arising as a result of repeated observations per area can be considered \citep{regression_fahrmeir}.\\

In general terms a mixed model is given by:
\begin{equation}
\label{grund_gleichung}
    \boldsymbol{y} = f(\boldsymbol{X}) + \boldsymbol{Z} \boldsymbol{\vartheta} + \boldsymbol{\epsilon}
\end{equation}

In the context of mixed model area modeling this results in

\begin{equation}
\label{grund_gleichung_2}
    \boldsymbol{y}_{d} = f(\boldsymbol{X}_{d}) + \boldsymbol{Z}_d \boldsymbol{\vartheta}_d + \boldsymbol{\epsilon}_d, \quad \quad d = 1,...,D ,
\end{equation}
for each domain $d$.\\
$ \boldsymbol{y}_{d}$ is an $n_d$-dimensional vector of objective variables of area $d$ with sample sizes $n_d$. $D$ defines the number of areas, so $N = \sum_{d = 1}^D n_d$. $\boldsymbol{X}_d$ and $\boldsymbol{Z}_d$ are ($n_d \times p$) and ($n_d \times q$)-dimensional design matrices to $p$ known covariates, where $q$ is limited to $q = 1$ in this setting, which means only one random effect exists per area. ${\boldsymbol{\vartheta}_d}$ is a d-dimensional vector of area-specific effects and $\boldsymbol{\epsilon}_d$ is a $q$-dimensional error vector \citep{regression_fahrmeir}. The distributional assumptions for $\boldsymbol{y}_d$ and $\boldsymbol{\epsilon}_d$ are:

\begin{equation*}
\begin{split}
    & {\boldsymbol{\vartheta}_d} \overset{iid}{\sim} N(0, \boldsymbol{G}_d)\\ 
    & {\boldsymbol{\epsilon}_d} \overset{iid}{\sim}  N(0, \boldsymbol{R}_d)\\ 
    & {\boldsymbol{\vartheta}} \ci {\boldsymbol{\epsilon}} 
\end{split}
\end{equation*}

$Cov(\boldsymbol{y}) = \boldsymbol{V} = diag(V_1, ..., V_d, ...., V_D)$, with $\boldsymbol{V}_d = \boldsymbol{R}_d + \boldsymbol{Z}_d \boldsymbol{G}_d \boldsymbol{Z}_d' $ defining the covariance matrix of the observations $\boldsymbol{y}$.
It also holds that:

\begin{equation*}
    \boldsymbol{Z} = diag(Z_1,...,Z_d,...,Z_D), \\ 
\end{equation*}
\begin{equation*}
 \boldsymbol{R} = diag(R_1, ..., R_d, ...., R_D) = \mathbb{1}_D\sigma_{\epsilon}^2  \text{ and } \boldsymbol{G} = diag(G_1, ..., G_d, ...., G_D) = \mathbb{1}_D\sigma_{\vartheta}^2,
\end{equation*} 
where $\mathbb{1}_D$ is the $D \times D$ identity matrix. From $\boldsymbol{R} = diag(R_1, ..., R_d, ...., R_D) = \mathbb{1}_D\sigma_{\epsilon}^2$ it can be deduced that only between-area variations are considered in this paper.
Setting $f(\boldsymbol{X})$ to a linear model $f(\boldsymbol{X}) = \boldsymbol{X}\boldsymbol{\beta}$ with regression parameters $\boldsymbol{\beta} = (\beta_1, ..., \beta_p)'$ as a special case of Equation \ref{grund_gleichung} results in the LMM proposed by \cite{battese}. Defining $f(\cdot)$ as a Random Forest results in the MERF approach proposed by \cite{MERF}, which \cite{Krennmair2022} extended to a small area context. Replacing the Random Forest used in the approach of \citeauthor{MERF} with a Gradient Boosting algorithm \citep{friedman_gb} results in the method introduced in Section \ref{Mixed effects gradient boosting}. In this paper, the used Gradient Boosting algorithm is the Extreme Gradient Boosting algorithm, proposed by \citep{xgboost_paper}. 

\subsection{Extreme Gradient Boosting}
Gradient Boosting refers to a class of ensemble machine learning algorithms that are constructed from decision tree models. The trees are added to the ensemble one at a time and adjusted to correct the prediction errors of the previous models. This is a type of ensemble model for machine learning called boosting, first presented by \cite{boosting_idea}. The idea of ARCing (Adaptive Reweighting and Combining) algorithms \citep{breiman_predictiongames} was extended by \cite{friedman_gb}. 
The models are fitted with an arbitrary differentiable loss function and a gradient descent optimization algorithm. This is where the name Gradient Boosting is derived from, as the loss gradient is minimized when fitting the model. A requirement for a loss function is that it must be differentiable \citep{friedman_gb}, in regression cases, this can be the squared error for example.
In this context, Gradient Boosting specifically employs decision trees as the foundational weak learners. Henceforth, any reference to Gradient Boosting within this paper pertains exclusively to Gradient Boosting with decision trees. A Gradient Boosting model is a series of weak decision trees that are created one after the other to form a strong learner. In addition, each weak decision tree is trained to correct the mistakes made by the previous weak learner decision trees.
The outcomes produced by the newly formed tree, when scaled by the learning rate $\eta$, are subsequently integrated with the outcomes from the sequence of existing trees. This process aims to refine or enhance the overall results generated by the model.
A fixed number of trees is added or training is terminated when the loss reaches an acceptable level or no longer improves. Extreme Gradient Boosting (XGBoost \citep{xgboost_paket}) is an efficient and effective implementation of the Gradient Boosting algorithm provided in an open-source library. 
It is an optimized implementation of Gradient Boosting, which enhances speed, performance and scalability through algorithmic improvements and system optimizations, including parallel processing and regularization features.
One of the main advantages of Gradient Boosting models is the absence of assumptions such as linearity or the distribution specification of model errors, but the observations are still assumed to be independent. 

As the performance of an XGBoost model largely depends on hyperparameters, they need to be tuned and adjusted to find a good trade-off between sufficient model complexity and the likelihood of encountering overfitting. Details about the procedure used in this paper can be found in the simulation sections \ref{mb} and \ref{db}.

\subsection{Mixed Effects Gradient Boosting}
\label{Mixed effects gradient boosting}
Combining Mixed Effects and Gradient Boosting results in the MEGB.
As mentioned in Section \ref{mixed_effects}, this approach is based on the Mixed Effect Random Forests of \cite{MERF} and the SAE extension of \cite{Krennmair2022}. The objective is to predict values for Out-Of-Sample observations, for which additional covariates from census or register information across areas are available, by identifying the relation $f(\cdot)$ as well as the random effects between covariate $\boldsymbol{X}$ and objective $\boldsymbol{y}$. 
The model is formed by a fixed effect part $f(\boldsymbol{X})$ and a random part $\boldsymbol{Z} \boldsymbol{\vartheta} $, which covers the dependencies introduced by the random effects \citep{Krennmair2022}.
The fixed part $f(\boldsymbol{X})$ of Equation \ref{grund_gleichung} is replaced by a Gradient Boosting Decision Tree model.

The assumption is made, that the distribution of the unit-level errors follows a normal distribution, given by $ \epsilon_i \sim  N(0, G)$ for unit $i$. In this context, the assumption of normality does not impact the efficacy of Gradient Boosting, and there is no requirement for the residuals to follow a normal distribution for the method to be applicable. As shown below, the Mixed Effect Gradient Boosting algorithm is estimated by an EM algorithm \citep{em_autoren}. 

However, regarding the part involving random components, an appropriate likelihood function is necessary. This is to guarantee that the EM algorithm, defined further below, successfully converges to a local maximum within the parameter space for estimating parameters \citep{Krennmair2022}.

\citeauthor{Krennmair2022}'s (\citeyear{Krennmair2022}) normality-based likelihood function is used because of two key benefits. Initially, it provides a closed-form solution of the integral of the Gaussian likelihood function, which facilitates the estimation of random effects. Additionally, by the mean of the unit-level residual sum of squares, the maximum likelihood estimate for the variance of the unit-level errors $\hat{\sigma}_{\epsilon}$ is given.

The EM algorithm to fit Equation \ref{grund_gleichung} is structured as follows: At the initial stage, the algorithm computes the fixed component of the response variable $y^*$. The currently estimated value of the corresponding random component $\vartheta$ is excluded by a Gradient Boosting algorithm. The number of trees of the Gradient Boosting algorithm, used at this stage in line 7 of Algorithm \ref{algorithm1}, follows an early stop criterion. Early stopping is applied when the model’s performance, measured by the root mean squared error (RMSE) on a cross-validation test subsample, does not improve for a specified number of rounds, halting further tree construction.
In the second phase, the algorithm deducts the estimated fixed component from the raw data $y$ to get a $y^{**}$ reduced by the fixed effect and an LMM is estimated. Subsequently, the random component is newly estimated.
The process continues iteratively through these steps until convergence is reached.  The algorithm enhances the accuracy of predictions by adjusting the fixed component and variance components based on the residuals after the fixed component has been accounted for, thus refining the model with each iteration to prevent overfitting. 
In summary, the MEGB algorithm first estimates the Gradient Boosting function, assuming
that the random effects term is correct. As a second step, it estimates the random effects part assuming that the predictions from the Gradient Boosting model are correct. The proposed algorithm for the Mixed Effect Gradient Boosting is shown below in Algorithm \ref{algorithm1}:\\

\begin{algorithm}
\caption{Mixed Effect Gradient Boosting}
\begin{algorithmic}[1]

\State \textbf{Initialization:}
\State set $\vartheta_{0} = 0$, \quad $\beta_0 = 0$
\State Set hyperparameters for Gradient Boosting (fixed or by tuning).

\State \textbf{EM Algorithm:}
\For{$iter = 1$ to $Iter_{max}$}
    \State Set $iter = iter + 1$.
    \State Subtract random effects $\vartheta_{iter}$ and intercept $\beta_0$ from the raw data $y$: $y^* = y - \beta_0 - Z\hat{\vartheta}_{iter-1}$.
    \State Fit a Gradient Boosting on the new corrected $y^*$ to obtain $\hat{f}_{iter}(y^*|\boldsymbol{X})$.
    \State Determine $y^{**} = y - \hat{f}_{iter}(y^*|\boldsymbol{X})$, the raw data $y$ subtracted by the GB estimate.
\State Estimate linear mixed model with intercept: $LMM(y^{**} | Z)$.
\State Estimate random effects $\hat{\vartheta}_{iter}$ and the intercept $\beta_0$.
\State $\left( \left|\frac{\log(\text{likelihood}_{iter}) - \log(\text{likelihood}_{{iter}-1})}{\log(\text{likelihood}_{{iter}-1})}\right| < \text{error tolerance}\right) \rightarrow$ Stop
\EndFor 

\end{algorithmic}
\label{algorithm1}
\end{algorithm}
The algorithm's convergence is evaluated by monitoring the marginal change in the modified Generalized Log-Likelihood (GLL) criterion at each iteration.
The GLL is given by \citep{MERF}

\begin{equation}
\begin{split}
    \text{GLL}(f,\vartheta_d| y) = & \sum_{i = 1}^n ( [y_i - f(X_i) - Z_i\vartheta_i]^T \cdot \\ & R_i^{-1} [y_i - f(X_i) - Z_i \vartheta_i] + \\ & \vartheta_i^T G^{-1} \vartheta_i + log|G| + log|R_i| ). 
\end{split}
\end{equation}
In the context provided, the index \( i \) is used to refer to unit \( i \), rather than to an area. \cite{Krennmair2022} derived the estimator for $\vartheta$ for known parameters $G$ and $R$ as $\hat{\vartheta} = GZ^TG^{-1}(y-\hat{f}(X))$. After convergence of the EM algorithm, the maximum likelihood estimators, given the data, deliver variance components $\hat{\sigma}_{\epsilon}^2$ and $\hat{\sigma}_{\vartheta}^2$ which are required for the uncertainty estimations described in Section \ref{uncertainty}.

\subsection{Small Area Average Predictions}
\label{small area average}
In the context of SAE, predictions on the individual level are of less interest than area level predictions \citep{Morales2021SmallAreaEstimation}.
Therefore, indicators like area-level means or area-level totals are needed. The MEGB model from Equation \ref{grund_gleichung_2} predicts conditional means on the individual level of a metric dependent variable $y_{di}$ given unit-level auxiliary information $x_{di}$. To align this approach with the outlined objectives, a method is needed, which generates area-level means or totals from individual predictions.

To facilitate this area-level estimation, we adopt structural simplifications similar to those used in the linear mixed models (LMM) suggested by \cite{battese}. This implies the existence of a single random effect, consequently resulting in the design matrix \(Z\) having the dimension \(n_i \times D\). The design matrix $Z$ is filled with area-intercept indicators, employing a \(D \times 1\) vector of random effects denoted as \(\vartheta = [\vartheta_1, \ldots , \vartheta_D]'\). This reduces the variance-covariance matrix for random effects to \(G_d = \mathbb{1}_D \sigma^2_{\vartheta}\) \citep{Morales2021SmallAreaEstimation}. $\mathbb{1}_D$ is the $D \times D$ identity matrix, which implies the independence of the $\vartheta_d$'s. 
To derive an estimator for area averages or totals, the unit-level \(y_i\)'s of the population data are estimated through the Gradient Boosting algorithm \(\hat{f}(X)\), employing supplementary data sources typically obtained from a census. The estimated \(\hat{y}_{d,i}\) are subsequently aggregated based on their respective area, with the addition of the estimated area-specific random effect \(\hat{\vartheta}_d\), leading to the estimated area mean $\hat{\mu}_d^{MEGB}$. The calculation of the area total $\hat{Y}_d^{MEGB}$ estimation follows the same procedure, with the distinction that it is not divided by the number $N_d$, which is also obtained from the census. 

\begin{equation}
  \hat{\mu}_d^{MEGB} = \overline{\widehat{f}}(x_d) + \hat{\vartheta}_d = N_d^{-1}\sum_{i=1}^{N_d} (\hat{f}(x_{d, i})) + \hat{\vartheta}_d
\end{equation}

\begin{equation}
  \hat{Y}_d^{MEGB} = \sum_{i=1}^{N_d} (\hat{f}(x_{d, i})) + N_d \cdot \hat{\vartheta}_d
\end{equation}

For areas that are not sampled, the parameter \(\hat{\vartheta}_d\) is omitted, and for the area-level mean estimation, the aggregated Gradient Boosting estimations, denoted as \(\overline{\widehat{f}}(x_d)\), builds the estimator for $\hat{\mu}_d^{MEGB}$. This approach allows for the efficient estimation of area-level means, aligning with the objectives of SAE.

\section{Uncertainty Estimation}
\label{uncertainty}

MSE estimates are an important component of SAE. In the case of the MERF, \cite{Krennmair2022} proposed a non-parametric bootstrap procedure based on the approach outlined in \cite{chambers_bootstrap}. This Random Effect Block  Bootstrap (REBB) procedure approach is adapted and applied to quantify the uncertainty for the MEGB method proposed in this paper. 

Block bootstrapping allows to expand bootstrapping to clustered data.
The aim is to preserve the cluster structure of the observations.
Keeping the cluster structure enables the quantification of uncertainty within the clusters \citep{carlstein1998matched}.

\cite{chambers_bootstrap} enhance block bootstrapping to especially account for random effects and variance within the blocks (level 1 variance), this is called Random Effect Block Bootstrapping. This requires that the level 1 variance is similar across the blocks. Algorithm \ref{rebb_algorithmus} shows the REBB used for the MEGB mean estimations. The REBB simulates the random effects for every bootstrap sample, in order to replicate the variance within clusters (blocks). This is done to model and quantify the uncertainty from within the clusters. First, using the marginal residuals, the average residuals for every cluster (level 2) and the residuals within the cluster (level 1) are calculated. Next, these residuals are scaled to the estimated variance ${\hat{\sigma}_{\epsilon}^2}$ and then centered. Respectively, the level 2 residuals are scaled to $\hat{\sigma}_{\vartheta}^2$ and then centered. In a next step, the scaled and centered residuals are used to generate independent bootstrap samples for the errors on level 2 and level 1. These errors are then used to simulate bootstrap data and then the Mixed Effect Gradient Boosting model is estimated on the bootstrap data. The proposed REBB has, in the case described here, the advantage of not relying on distributional assumptions.   
The clusters in this context refer to the domains in SAE. 
We denote the outcome of selecting a simple random sample of size $D$ with replacement from set $A$ as $srswr(A, D)$. Algorithm \ref{rebb_algorithmus} is used to calculate the area mean residuals but can easily be adapted to calculate the total area residuals, by following the same procedure as for the area mean residuals, with the exception that the residuals are not divided by $n_d$.

\begin{algorithm}[htb!]
    \caption{Random Effect Block Bootstrap}
    \begin{algorithmic}[1]
      \State \textbf{Calculate residuals:}
        \Statex Calculate the marginal residuals within each area $d$ and observation $i$ 
        \begin{equation*}
           \hat{e}_{d,i} = y_{d,i} - \hat{f}_{MEGB}(X_{d,i})
        \end{equation*} 
        for a given $\hat{f}(\cdot)$. 
        \Statex Calculate the level 2 average residuals for each of the $D$ areas using the level 1 residuals:
        \begin{equation*}
            \bar{r}_d^{(2)} = n_{d}^{-1} \sum_{i = 1}^{n_d} \hat{e}_{d,i}
        \end{equation*}

        \Statex Calculate the level 1 residuals for each area $d$: $\hat{r}_{d,i}^{(1)}  = \hat{e}_{d,i} - \bar{r}_d^{(2)}, \quad i = 1, \ldots, n_d, \quad d = 1, \ldots, D$
        \Statex $\hat{r}_{d,i}^{(1)}$ are scaled to the estimated variance $\hat{\sigma}_{\epsilon}^2$ and subsequently centered. This results in the scaled and centered level 1 residuals $\hat{r}_{d,i}^{1,c}$. Analogously, level 2 residuals $\bar{r}_d$ are scaled to $\hat{\sigma}_{\vartheta}^2$ and centered to yield $\bar{r}_d^{2,c}$, mirroring the process for level 1.
        \Statex $\Bar{\bold{{r}}}^{2,c}$ is, in this case, the $D \times 1$ scaled and centered vector of the $D$ area level (level 2) average residuals. $\bar{r}_d^{2,c}$ and $\hat{\bold{r}}_d^{1,c}$ is the $n_d \times 1$ scaled and centered vector of unit level (level 1) residuals $\hat{r}_{d,i}^{1, c}$ for group $d$. 
      \State \textbf{Definition of bootstrap sample errors}:
        \Statex Sample independently with replacement from the two sets of centred and scaled residuals. The level 2 bootstrap errors are given by $\bar{\bold{r}}^{2,b} = srswr(\Bar{\bold{r}}^{2,c}, D$), whereas the level 1 bootstrap errors are given by $\hat{\bold{r}}_{d}^{1,b} = srswr(\hat{\bold{r}}_{d,i}^{1, c}, n_d)$. 
    \State \textbf{Simulate bootstrap data}:
    \Statex Use the model
    \begin{equation*}
        y_{d,i}^{(b)} = \hat{f}(X_{d,i}) + Z\bar{r}_d^{2,b} + \hat{r}_{d, i}^{1,b} 
    \end{equation*}
    to simulate the bootstrap population data. 
    \Statex Calculate the bootstrap population area means $\mu_d^{(b)} = N_d^{-1} \sum_{i=1}^{N_d} y_{d, i}^{(b)}$.
    \Statex Draw bootstrap sample data from the bootstrap population maintaining sample size $n_d$ from the original sample. 
    \State \textbf{Fit the Mixed Effect Gradient Boosting algorithm}:
    \Statex To get the bootstrap area level estimates $\hat{\mu}_d^{(b)}$, fit the MEGB model (Section \ref{Mixed effects gradient boosting}) with the hyperparameters of the point estimation to the bootstrap sample data generated in the step before and estimate the area mean (Section \ref{small area average}).
    \State \textbf{Repeat B times for B sets of bootstrap parameter estimates}:
    \Statex Repeat the steps $2-4$ $B$ times to generate $B$ area mean estimates. 
    \end{algorithmic}
    \label{rebb_algorithmus}
\end{algorithm}

The estimation of the Mean Squared Error for domain $d$ is given by:

\begin{equation}
    \widehat {MSE}_d = B^{-1} \sum_{b = 1}^{B} (\mu_{d}^{(b)} -\hat{\mu}_{d}^{(b)})^2 
\end{equation}

\section{Model-based Simulations}
\label{mb}
\subsection{Simulation Setup, Comparative Estimators, Evaluation Metrics}
\label{mb_setup}
The MEGB's performance is tested through simulation studies in four different scenarios, and compared to different estimation methods; i.e. the BHF implemented in the sae-package \citep{sae} in R \citep{R}. In addition, the Empirical-Best-Prediction with box-cox transformation (EBP-BC) from the emdi-package \citep{emdi} and the Mixed Effect Random Forest (MERF) from the SAEforest-package \citep{SAEforest} are used for comparison. The four different simulation scenarios are set up as follows: Across all scenarios, the random effects $\vartheta$ are normally distributed with $\mu = 0$. Also, the covariates $x_1$ and $x_2$ are normally distributed. The Linear-Normal and the Complex-Normal scenario both have normally distributed residuals. However, while the Linear-Normal scenario uses a linear relationship, the Complex-Normal model is more elaborate, as it includes an interaction $x_1 \cdot x_2$ and a quadratic term $x_2^2$. The Linear-Pareto and Complex-Pareto scenarios are analogous to the Linear-Normal and Complex-Normal scenarios, with the difference that now, instead of the normal distribution, the Pareto distribution is used for the error term $\epsilon$. The Pareto distribution has in these cases the parameters $k = 3$ and $a = 800$. Therefore, the distribution is always positive with values above three and is skewed to the right. Using the Pareto distribution makes it possible to test the MEGB in a more complex scenario with error terms skewed to the right. The exact scenarios and models for each scenario can also be seen in Table \ref{scenarios} below. Each scenario is repeated for $200$ Monte-Carlo simulation runs. The setting for the simulations is as follows: It consists of a population $\bm{U}$ with size $N = 50 000$ and $D = 50$ disjunct areas $U_1, ..., U_D$ of size $N_d = 1000$ for all areas $d$. The sample size created through stratified random sampling is given as $n = \sum_{d=1}^D n_d = 1410$, where the observations vary between 6 and 49 per area, with $\bar{n}_d = 28$ per area. The simulation setting is the same as in the paper of \cite{Krennmair2022} and is used to compare the MEGB and the MERF as well as the settings used for the MERF.

The point estimates for the area means are evaluated using the relative root mean squared error (RRMSE) and the relative bias (RB) out of the $n_{MC}$ Monte-Carlo simulation runs. They are defined as follows: 

\begin{equation*}
    RB_d = n_{MC}^{-1}  \sum_{j = 1}^{n_{MC}} \frac{\hat{\overline{\mu}}_{d, j}- \mu_{d, j}}{\mu_{d, j}},
\end{equation*}

\begin{equation*}
    RRMSE_d = \frac{\sqrt{n_{MC}^{-1} \sum_{j = 1}^{n_{MC}}(\hat{\overline{\mu}}_{d, j}- \mu_{d, j})^2}}{n_{MC}^{-1} \sum_{j = 1}^{n_{MC}}\mu_{d, j}} ,
\end{equation*}

\begin{equation*}
    RB-RMSE_d = \frac{\sqrt{n_{MC}^{-1} \sum_{j = 1}^{n_{MC}}MSE^{(est)}_{d,j}}-RMSE_{d}^{(emp)}}{RMSE_{d}^{(emp)}} \cdot 100 ,
\end{equation*}

\begin{equation*}
    RRMSE-RMSE_d = \frac{\sqrt{n_{MC}^{-1}\sum_{j = 1}^{n_{MC}}(\sqrt{MSE^{(est)}_{d,j}}-RMSE_{d}^{(emp)})^2}}{RMSE_{d}^{(emp)}} \cdot 100,
\end{equation*}

where $MSE^{(est)}_{d,j}$ is the estimated MSE for area $d$ in Monte-Carlo simulation run $j$ and
$\hat{\overline{\mu}}_{d, j}$ is the estimated area average for area $d$ in Monte-Carlo run $j$, where $j$ denotes one of $n_{MC}$ Monte-Carlo simulation runs.

To evaluate the MSE estimates, the relative bias and the relative root mean squared error of the MSE estimates are used. 
The estimation of the MSE follows the method proposed in Section \ref{uncertainty}, with $B = 100$. The MSE is then calculated using the already available empirical true values. In Section \ref{mb_results} the empirical true and the estimated MSE values are compared to assess the performance of the used MSE estimation bootstrap procedure.

The hyperparameters of the XGBoost model, except for the number of trees ($n_{rounds}$) are already chosen before running the EM algorithm and are therefore fixed. The configuration specifies a learning rate $\eta$ of $0.01$, controlling the step size for each iteration in the optimization process. The parameter \textit{max\_depth} is limited to $3$ and defines the maximum depth of each tree. The \textit{min\_child\_weight} is also set to $3$, affecting decisions on further splitting at the nodes. The subsample-parameter is adjusted to $0.5$, which specifies that only half of the data samples are used to grow the trees. Furthermore, two regularization parameters are applied through a value of $\lambda = 1$, which refers to the L2 regularization term on weights. Additionally, $\gamma$ is set to $0.9$, specifying the minimum loss reduction required to make a further partition. 

The results with tuned parameters using only 20 simulation runs and details on parameter tuning are shown in the Appendix. The parameter $n_{rounds}$ follows an early stop criterion in the case of tuned and fixed hyperparameters. 

\begin{table}[]
\resizebox{\textwidth}{!}{
\begin{tabular}{lllllll}
\hline
\multicolumn{1}{l}{Scenario} & \multicolumn{1}{l}{Model}                                          & \multicolumn{1}{l}{$x_1$} & \multicolumn{1}{l}{$x_2$} & \multicolumn{1}{l}{$\mu$} & \multicolumn{1}{l}{$\vartheta$} & \multicolumn{1}{l}{$\epsilon$} \\ \hline
Linear-Normal                         & $y = 5000 - 500x_1-500x_2 +\vartheta + \epsilon$               & $N(\mu, 3^2)$              & $N(\mu, 3^2)$                & $unif(-1,1)$               & $N(0, 500^2)$                    & $N(0, 1000^2)$                 \\
Complex-Normal             & $y = 15000 - 500x_1 \cdot x_2- 250x_2^2 +\vartheta + \epsilon$ & $N(\mu, 4^2)$              & $N(\mu, 2^2)$                & $unif(-1,1)$               & $N(0, 500^2)$                    & $N(0, 1000^2)$                 \\
Linear-Pareto                  & $y = 5000 - 500x_1-500x_2 +\vartheta + \epsilon$               & $N(\mu, 3^2)$              & $N(\mu, 3^2)$                & $unif(-1,1)$               & $N(0, 500^2)$                    & $Par(3,800)$                   \\
Complex-Pareto             & $y = 20000 - 500x_1 \cdot x_2-250x_2^2 +\vartheta + \epsilon$ & $N(\mu, 2^2)$              & $N(\mu, 2^2)$                & $unif(-1,1)$               & $N(0, 1000^2)$                    & $Par(3,800)$   \\ \hline

\end{tabular}
}
\caption{Scenarios used in the model-based simulation study}
\label{scenarios}
\end{table}
\subsection{Simulation Results}
\label{mb_results}
\begin{figure}[htb!]
    \centering
    \includegraphics[width = \linewidth]{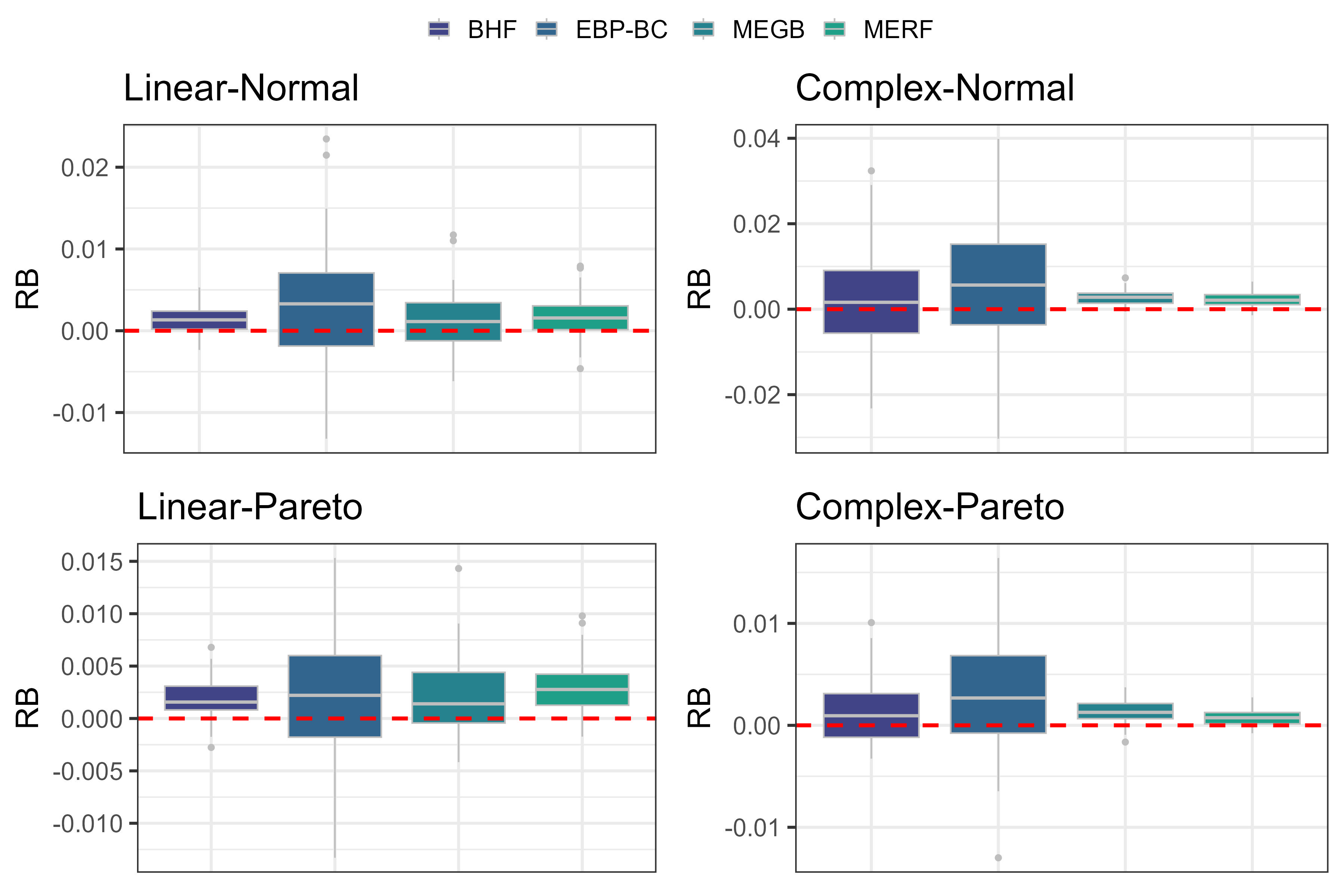}
    \caption{Relative bias averaged over the MC simulation runs for the four scenarios}
    \label{fig:mb_relative Bias}
\end{figure}

In this subsection, the results of the simulation described in Subsection \ref{mb_setup} will be discussed and evaluated using the relative bias (RB) and the relative root mean squared error (RRMSE). Figure \ref{fig:mb_relative Bias} shows boxplots of the relative bias for the four scenarios described in Table \ref{scenarios} and the corresponding mean and median relative bias values are available in Table \ref{rb mb}. 

\begin{table} [htb!]
\centering
\resizebox{\textwidth}{!}{ 
\begin{tabular}{lllllllll}
\toprule
\multicolumn{1}{c}{RB (\%)} & \multicolumn{2}{c}{Linear-Normal} & \multicolumn{2}{c}{Complex-Normal} & \multicolumn{2}{c}{Linear-Pareto} & \multicolumn{2}{c}{Complex-Pareto} \\
\cmidrule(l{3pt}r{3pt}){1-1} \cmidrule(l{3pt}r{3pt}){2-3} \cmidrule(l{3pt}r{3pt}){4-5} \cmidrule(l{3pt}r{3pt}){6-7} \cmidrule(l{3pt}r{3pt}){8-9}
  & Mean & Median & Mean & Median & Mean & Median & Mean & Median\\
\midrule
BHF & 0.133 & 0.133 & 0.239 & 0.161 & 0.190 & 0.157 & 0.127 & 0.094\\
EBP-BC & 0.279 & 0.329 & 0.576 & 0.565 & 0.190 & 0.221 & 0.308 & 0.268\\
MEGB & 0.127 & 0.113 & 0.260 & 0.280 & 0.207 & 0.140 & 0.134 & 0.127\\
MERF & 0.166 & 0.156 & 0.237 & 0.210 & 0.310 & 0.277 & 0.072 & 0.074\\
\bottomrule
\end{tabular}
}
\caption{Mean and the median relative bias in \% for the four estimators across the four scenarios}
\label{rb mb}
\end{table}

\begin{figure}[htb!]
    \centering
     \includegraphics[width = \linewidth]{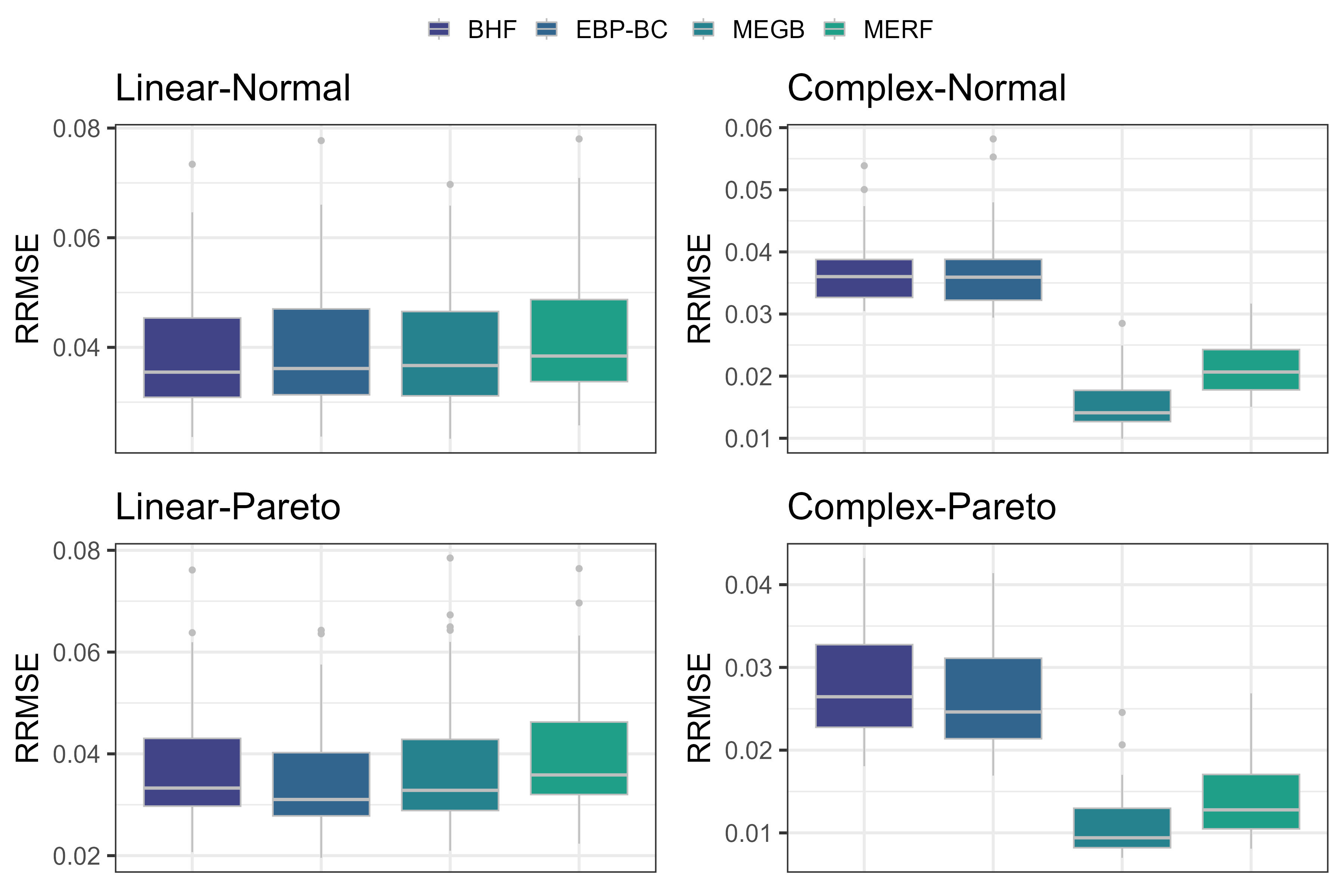}
    \caption{Relative RMSE averaged over the MC simulation runs for the four estimators across the four scenarios}
   \label{fig:mb_rrmse boxplot}
 \end{figure}

The results in Figure \ref{fig:mb_relative Bias} and Figure \ref{fig:mb_rrmse boxplot} show, that for a linear relationship with normally distributed residuals, the BHF and the EBP-BC seem to perform slightly better in terms of a lower RRMSE than the MERF and the MEGB. This is also underlined by Table \ref{tab: rrmse_tabelle_mb}, where BHF and EBP-BC have a median RRMSE of 3.55\% and 3.61\% as opposed to the higher mean RRMSE of 3.84\% for the MERF and 3.67\% for the MEGB in the Linear-Normal scenario. One would expect the BHF and EBP-BC to perform best in the Linear-Normal scenario. In this scenario, the BHF has the best results, which aligns with the expectations made beforehand, as the data generating process fits the model of the EBP-BC and the BHF.

\begin{table}[htb!]
\centering
\resizebox{\textwidth}{!}{ 
\begin{tabular}{lllllllll}
\toprule
\multicolumn{1}{c}{RRMSE (\%)} & \multicolumn{2}{c}{Linear-Normal} & \multicolumn{2}{c}{Complex-Normal} & \multicolumn{2}{c}{Linear-Pareto} & \multicolumn{2}{c}{Complex-Pareto} \\
\cmidrule(l{3pt}r{3pt}){1-1} \cmidrule(l{3pt}r{3pt}){2-3} \cmidrule(l{3pt}r{3pt}){4-5} \cmidrule(l{3pt}r{3pt}){6-7} \cmidrule(l{3pt}r{3pt}){8-9}
  & Mean & Median & Mean & Median & Mean & Median & Mean & Median\\
\midrule
BHF & 3.907 & 3.548 & 3.652 & 3.602 & 3.770 & 3.328 & 2.755 & 2.645\\
EBP-BC & 3.986 & 3.614 & 3.660 & 3.593 & 3.512 & 3.103 & 2.630 & 2.462\\
MEGB & 3.982 & 3.669 & 1.544 & 1.410 & 3.768 & 3.284 & 1.088 & 0.941\\
MERF & 4.187 & 3.842 & 2.171 & 2.066 & 4.021 & 3.586 & 1.372 & 1.278\\
\bottomrule
\end{tabular}
}
\caption{Mean and the median Relative RMSE in \% for the four estimators across the four scenarios}
\label{tab: rrmse_tabelle_mb}
\end{table}

The Linear-Pareto scenario - which consists of a linear relationship, yet in contrast to the Linear-Normal scenario the residuals are not normally distributed - shows slight changes in the performances of the estimators compared to the Linear-Normal scenario. The estimation quality of the MEGB is in this case similar to the estimation quality of the BHF, while the EBP-BC seems to obtain the best results in this scenario. This can be seen in Figure \ref{fig:mb_rrmse boxplot} as well as in Table \ref{tab: rrmse_tabelle_mb}. This result was to be expected, as the EBP is used with the box-cox transformation. 
The EBP-BC achieves in this scenario a mean RRMSE of $3.51\%$, while MEGB and BHF have a very similar mean RRMSE with $3.768\%$ and $3.770\%$ respectively, as is shown in in Table \ref{tab: rrmse_tabelle_mb}. Furthermore, it is shown in Figure \ref{fig:mb_rrmse boxplot} that the Mixed Effect Random Forest performs slightly worse than the other estimators in this scenario.

The RRMSE seems to follow a similar pattern across the four estimators in the case of the Complex-Normal and Complex-Pareto scenarios. The BHF and EBP-BC are outperformed by MERF and MEGB in both scenarios. MERF and MEGB seem to have a higher estimation quality than BHF and EBP, in addition, the MEGB has a lower RRMSE than the MERF; in the case of the Complex-Pareto scenario, the MEGB has a mean RRMSE of $1.09\%$ as opposed to the MERF's mean RRMSE of $1.37\%$.
Again, this result matches the expectations made beforehand, as one strength of the proposed MEGB is that complex relationships are detected automatically and therefore have no negative impact on the performance of the MEGB. Using residuals, which are not normally distributed, seems to have no effect on the estimation quality of the MERF or the MEGB, as can be seen in the scenarios using Pareto distributed error terms.


\begin{figure}[htb!]
    \centering
    \includegraphics[width = \linewidth]{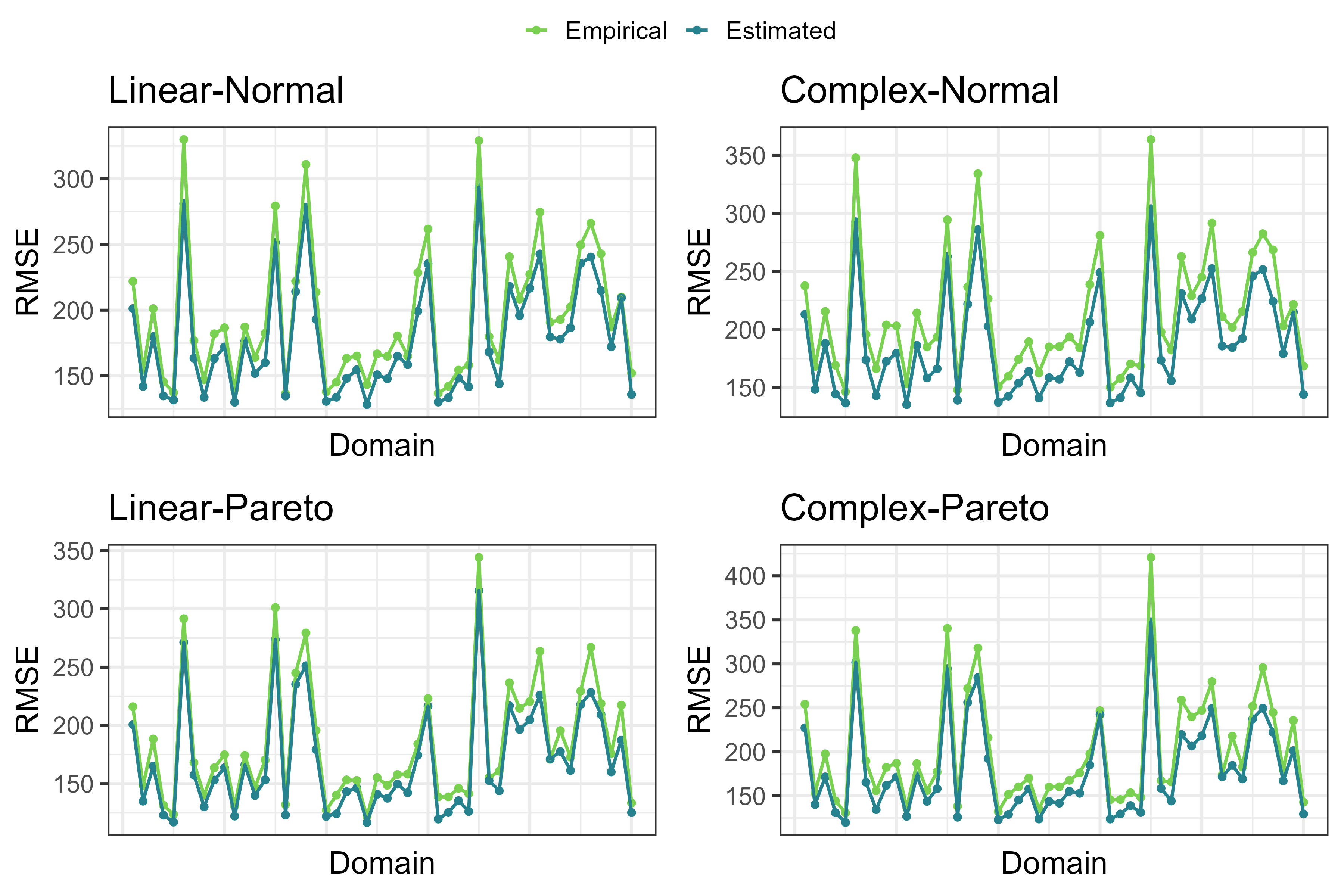}
    \caption{Empirical bootstrapped RMSE compared to the true RMSE across all four scenarios}
    \label{fig:mb_mse_est}
\end{figure}

\vspace{1cm}

Table \ref{tab:mb rb-rmse und rrmse-rmse} and Figure \ref{fig:mb_mse_est} can be used to assess the quality of the proposed bootstrap MSE estimation model.
Figure \ref{fig:mb_mse_est} shows the empirical RMSE as well as the RMSE estimated with the described bootstrap MSE estimation method across all four scenarios. The results from Table \ref{tab:mb rb-rmse und rrmse-rmse} and Figure \ref{fig:mb_mse_est} indicate that the proposed MSE estimation procedure is able to handle different levels of complexity. The estimated values seem to match and track the empirical RMSE in all scenarios quite well and no systematic difference is observed.

\begin{table}[htb!]
\centering
\resizebox{\textwidth}{!}{ 
\begin{tabular}{lrrrrrrrr}
\toprule
\multicolumn{1}{c}{ } & \multicolumn{2}{c}{Linear-Normal} & \multicolumn{2}{c}{Complex-Normal} & \multicolumn{2}{c}{Linear-Pareto} & \multicolumn{2}{c}{Complex-Pareto} \\
\cmidrule(l{3pt}r{3pt}){2-3} \cmidrule(l{3pt}r{3pt}){4-5} \cmidrule(l{3pt}r{3pt}){6-7} \cmidrule(l{3pt}r{3pt}){8-9}
  & Mean & Median & Mean & Median & Mean & Median & Mean & Median\\
\midrule
RB-RMSE & -0.083 & -0.080 & -0.123 & -0.108 & -0.074 & -0.064 & -0.105 & -0.099\\
RRMSE-RMSE & 15.780 & 16.097 & 17.139 & 17.130 & 40.457 & 43.217 & 36.517 & 39.806\\
\bottomrule
\end{tabular}
}
\caption{Mean and median results of the proposed bootstrap via RB-RMSE and RRMSE-RMSE for the four scenarios}
\label{tab:mb rb-rmse und rrmse-rmse}
\end{table}
The results show the advantages of combining Gradient Boosting with Mixed Effects in a SAE context.

\section{Design-based Simulation}
\label{db}

\subsection{Simulation Setup}
 In this section, the MEGB's performance is tested and evaluated using real data. The proposed method is used to estimate the area-level average labour income of Mexican municipalities in Nuevo Léon. The used data is a combination of a sample of census micro data with an income and expenditure survey of households. The objective is to model the per capita income obtained from employment. The earned income per capita is available in the survey data, as well as in the used census data. 
 The census dataset consists of $N = 54 848$ households from all $D =  51$ municipalities, while the survey data consists of $n = 1435$ households from 21 municipalities, with an average of $\bar{n}_d = 28$ households per municipality and 30 Out-Of-Sample municipalities. It holds that $min(n_d) = 5$ and $max(n_d) = 342$. The dataset used in this application is the same as the one used in \cite{Krennmair2022}. The estimation is repeated for $100$ Monte-Carlo simulation runs. Here, the hyperparameter \textit{colsample\_bytree} is used additionally to the hyperparameters already used in the model-based simulation, because more covariates are included in this dataset. The hyperparameter \textit{colsample\_bytree} impacts the subsample ratio of columns when each tree is constructed and is set to $0.6$ in this context. Analogously to the model-based settings, the hyperparameters were chosen before the analysis and are treated as fixed, except for the hyperparameter $n_{rounds}$, which again is set by the early stopping criterion. Similar to the model-based simulation, results with tuned parameters using only 50 simulation runs can be found in the Appendix.

\subsection{Results}
\label{db_results}
The bias and the empirical RMSE are used to assess the performance of the MEGB in relation to the other estimators (BHF, EBP-BC, and MERF). The Horvitz-Thompson estimator \citep{horvitz1952generalization}, denoted as HT, is also included in the analysis as a direct estimator. 

The HT mean estimator is denoted as $\hat{\mu}^{HT}_d = n_d^{-1} \sum_{i=1}^{n_d} \frac{y_{d,i}}{\pi_{d,i}}$, where $\pi_{d,i}$ is the probability that unit $d,i$ is included in the sample. \\

It should be noted, that the HT-estimator is only included for the In-Sample data, which consists of 21 municipalities. This limitation arises because the Horvitz-Thompson (HT) estimator is applicable only when the data points, represented by \( y_{d,i} \), are directly accessible for the estimation process. The empirical RMSE is calculated using the empirical true values, where the empirical true values are defined as the area-level averages of household incomes from labour in the used census data. The design-based simulation  also enables the analysis of Out-Of-Sample data, in addition to the In-Sample data.

\begin{figure}[htb!]
    \centering
    \includegraphics[width = \linewidth]{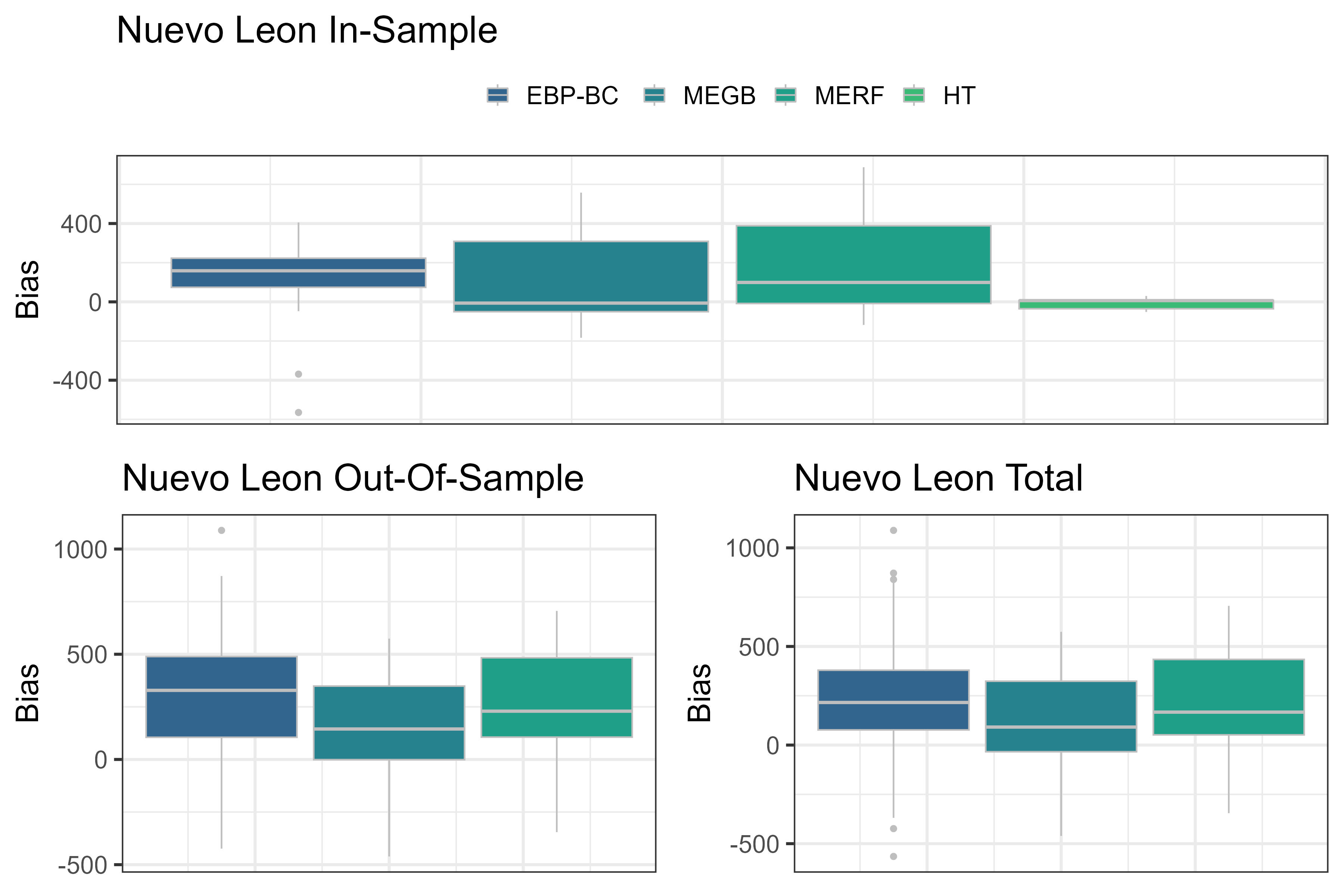}
    \caption{Bias for In-Sample -, Out-Of-Sample- and Total municipalities in Nuevo Léon for the EBP-BC, MEGB, MERF and the direct estimator \cite{horvitz1952generalization}.}
    \label{db_rb}
\end{figure}

\begin{table}[htb!]
\centering
\begin{tabular}{lllllll}
\toprule
\multicolumn{1}{c}{Bias} & \multicolumn{2}{c}{In-Sample} & \multicolumn{2}{c}{Out-Of-Sample} & \multicolumn{2}{c}{Total} \\
\cmidrule(l{3pt}r{3pt}){1-1} \cmidrule(l{3pt}r{3pt}){2-3} \cmidrule(l{3pt}r{3pt}){4-5} \cmidrule(l{3pt}r{3pt}){6-7}
  & Mean & Median & Mean & Median & Mean & Median\\
\midrule
EBP-BC & 109.286 & 158.782 & 319.249 & 328.412 & 232.794 & 215.931\\
MEGB & 102.209 & -6.291 & 157.357 & 144.780 & 134.649 & 91.270\\
MERF & 188.521 & 99.188 & 255.364 & 229.452 & 227.841 & 167.026\\ 
\bottomrule
\end{tabular}
\caption{Mean and median bias for In-Sample-, Out-Of-Sample-, and all municipalities (Total) of the design-based simulation study}
\label{db_bias_fix}
\end{table}
The bias of the mean area level estimates is available in Figure \ref{db_rb}. In terms of a low bias, the MEGB outperforms the EBP-BC and MERF. This difference in estimation quality is underlined in Table \ref{db_bias_fix}. The MEGB has the lowest median and mean bias for In-Sample areas as well as for Out-Of-Sample areas.

\begin{figure}[htb!]
    \centering
    \includegraphics[width = \linewidth]{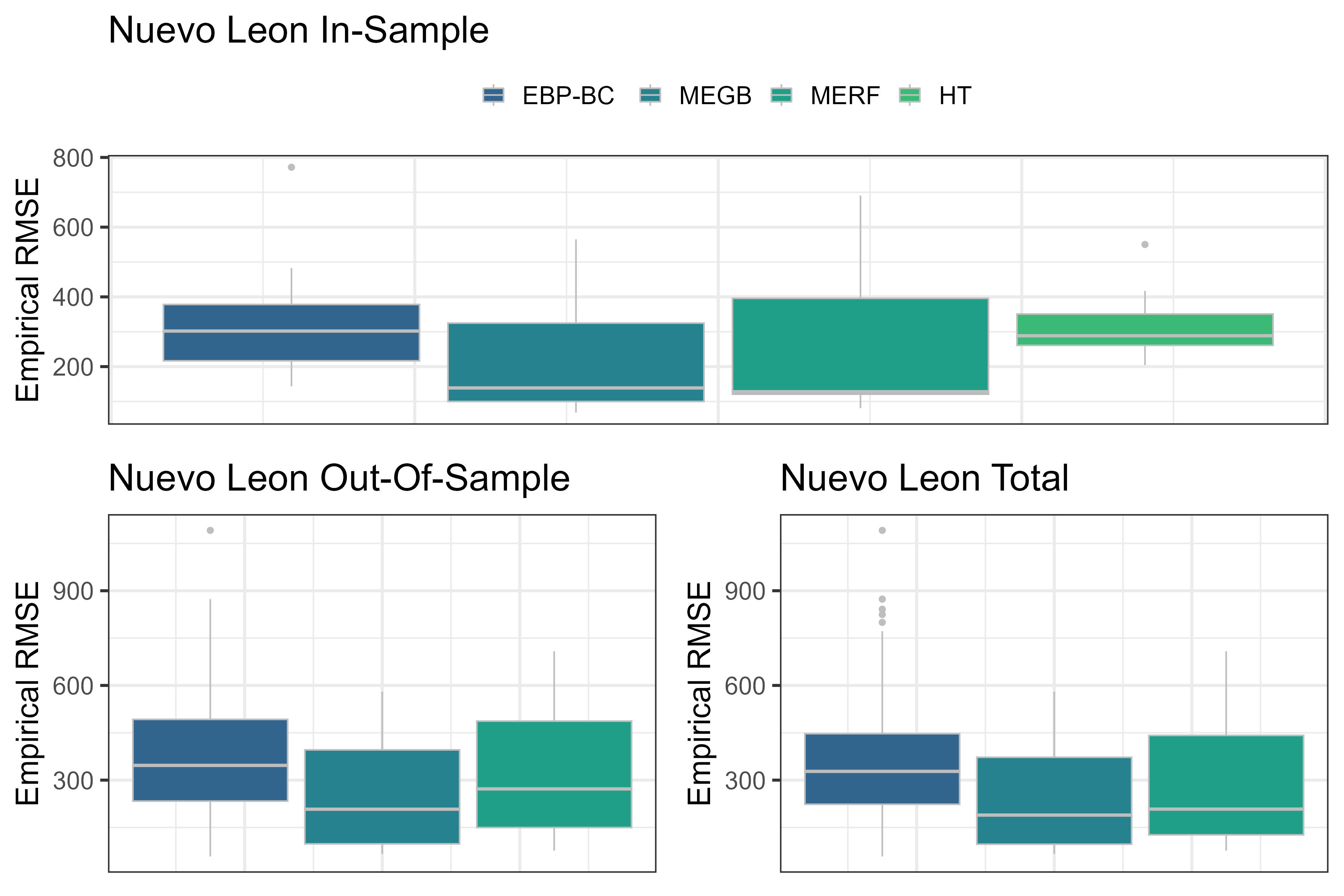}
    \caption{RMSE averaged over the MC simulation runs for the estimators EBP-BC, MEGB, MERF and the direct estimator}
    \label{fig:db_empirical rmse fix}
\end{figure}

To get a better impression of the MEGB compared to the EBP-BC and MERF, the empirical RMSE is also considered. The empirical RMSE of the mean area level estimates is presented in Figure \ref{fig:db_empirical rmse fix}. The MEGB has a lower empirical RMSE than the EBP-BC and the HT and is for the In-Sample data on a similar level as the MERF. Table \ref{db_rmse_fix} shows, that the MEGB performs also better than the MERF and has the lowest mean and median empirical RMSE across all data samples, except for the In-Sample median, where the MERF has a slightly lower median than the MEGB. This can be seen especially for the 30 Out-Of-Sample municipalities where the difference between the mean and median empirical RMSE for the MEGB and the other two estimators is the highest. The mean empirical RMSE is in this case $257.42$ for the MEGB, while MERF and EBP-BC have a mean empirical RMSE of $316.19$ and $417.59$ respectively. This indicates a good generalizability of the MEGB, which is an important feature for Out-Of-Sample estimations. 

\begin{table}[htb!]
\centering
\begin{tabular}{lllllll}
\toprule
\multicolumn{1}{c}{RMSE} & \multicolumn{2}{c}{In-Sample} & \multicolumn{2}{c}{Out-Of-Sample} & \multicolumn{2}{c}{Total} \\
\cmidrule(l{3pt}r{3pt}){1-1} \cmidrule(l{3pt}r{3pt}){2-3} \cmidrule(l{3pt}r{3pt}){4-5} \cmidrule(l{3pt}r{3pt}){6-7}
  & Mean & Median & Mean & Median & Mean & Median\\
\midrule
EBP-BC & 316.664 & 302.027 & 417.592 & 346.681 & 376.034 & 327.724\\
MEGB & 219.948 & 138.814 & 257.417 & 207.796 & 241.989 & 189.206\\
MERF & 256.574 & 127.962 & 316.186 & 272.137 & 291.640 & 208.215\\
\bottomrule
\end{tabular}
\caption{Mean and the median RMSE for the three estimators and the In-Sample, Out-Of-Sample and Total areas}
\label{db_rmse_fix}
\end{table}

\begin{figure}[htb!]
    \centering
    \includegraphics[width = \linewidth]{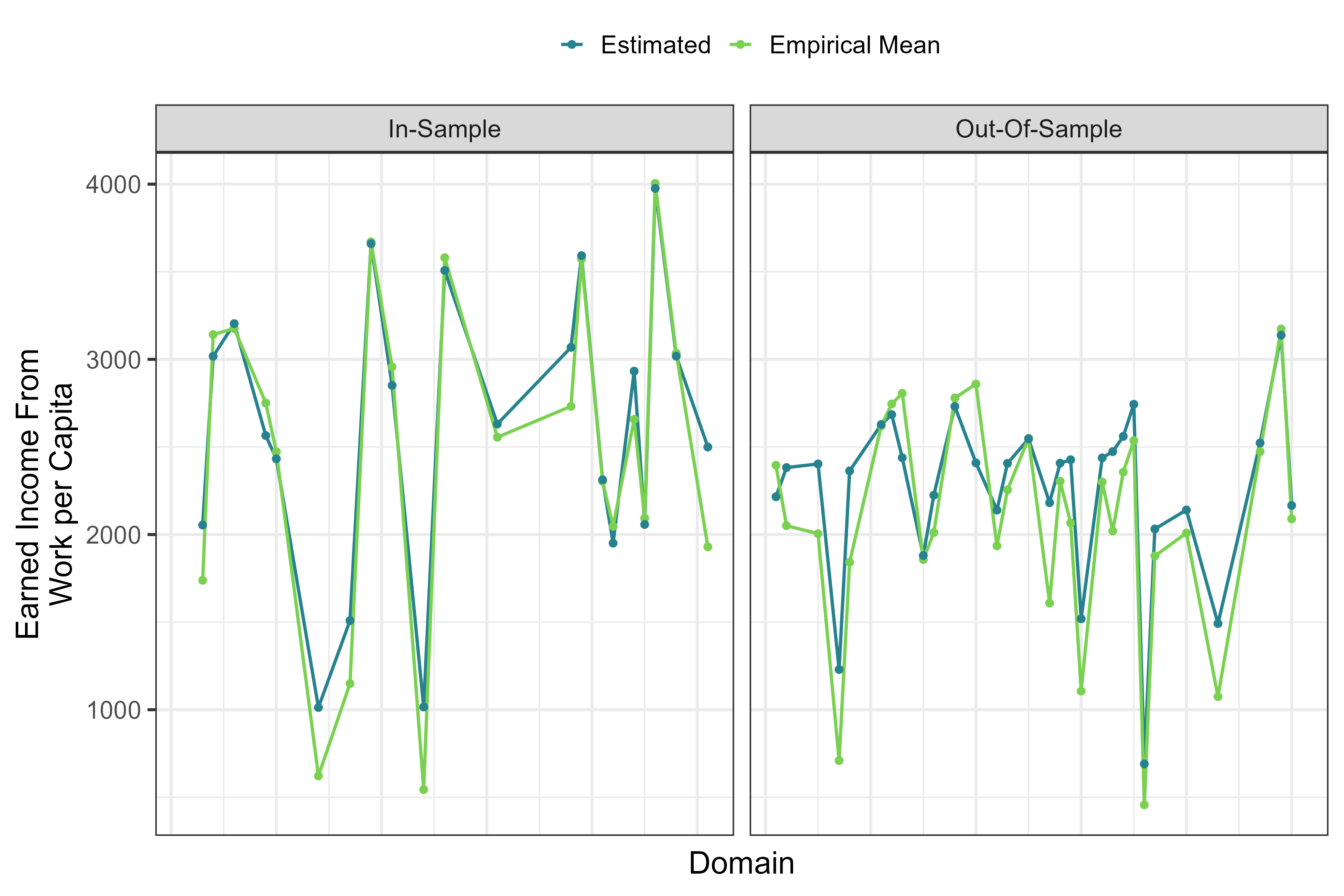}
    \caption{Point estimates of the average household income per area compared to the empirical true average income per area for In-Sample and Out-Of-Sample areas}
    \label{lineplot_db}
\end{figure}

The point estimates of the area means, which are shown in Figure \ref{lineplot_db}, track the empirical means well. This is especially the case for the In-Sample areas. In addition, it is displayed that the estimations for the Out-Of-Sample areas are close to the empirical values and provide correspondingly good area mean estimations, which highlights the generalizability and usability of the MEGB.

\begin{table}[htb!]
\centering
\begin{tabular}{lrrrrrr}
\toprule
\multicolumn{1}{c}{ } & \multicolumn{2}{c}{In-Sample} & \multicolumn{2}{c}{Out-Of-Sample} & \multicolumn{2}{c}{Total} \\
\cmidrule(l{3pt}r{3pt}){2-3} \cmidrule(l{3pt}r{3pt}){4-5} \cmidrule(l{3pt}r{3pt}){6-7}
  & Mean & Median & Mean & Median & Mean & Median\\
\midrule
RB-RMSE & 0.165 & 0.029 & 0.356 & 0.083 & 0.277 & 0.049\\
RRMSE-RMSE & 56.538 & 44.258 & 76.782 & 53.984 & 68.446 & 48.681\\
\bottomrule
\end{tabular}
\caption{Mean and median results of the proposed bootstrap through RB-RMSE and RRMSE-RMSE for the In-Sample, Out-Of-Sample and Total areas}
\label{db_rb_rmse}
\end{table}

The RB-RMSE and RRMSE-RMSE for the MEGB can be seen in Table \ref{db_rb_rmse} which allows the analysis of the proposed MSE estimation method. The table shows that the MSE estimations achieve lower results in In-Sample areas than in Out-Of-Sample areas. This aligns with the expectations made beforehand, as the estimation of the OOS areas is solely based on the Gradient Boosting algorithm $\hat{f}(x)_{MEGB}$ and no estimation of the random effect $\hat{\vartheta}_{OOS}$ can be performed.

\section{Conclusion and Future Research}
\label{conclusion}

Building on the estimation approach of the MERF, a combination of Random Forests and Mixed Effects, the aim was to develop an estimation method that meets the requirements of flexible, powerful machine learning methods combined with Mixed Effects. To achieve this, the method Mixed Effect Gradient Boosting was introduced. Combining the MEGB with Random Effect Block Bootstrapping creates a flexible estimator, which can be used without the constraints imposed by distributional assumptions. This flexibility distinguishes the MEGB from other estimation methods, as it automatically detects and exploits (higher-order) relations between predictive covariates. This feature significantly lowers the risk of model misspecification, as is the case with a few other estimators. Therefore, the MEGB method is also suitable for scenarios with high-dimensional problems processing big data sources and non-normally distributed residuals and yields accurate estimation results. The results of this paper show, that the combination of the mentioned Gradient Boosting model advantages together with the ability to control for dependency structures, builds a powerful method for SAE. 
Through the evaluation of the model-based scenarios and the design-based simulation, it was possible to show that the MEGB produced good point estimates for the area means across all scenarios. The MEGB outperformed the MERF in all four model-based and in the design-based scenarios.

Looking ahead, further investigation could research the extension of the proposed method to binary or count data, as well as examine the potential of the MEGB to estimate nonlinear indicators such as head-count ratios or Gini coefficients, as such indicators often hold significant relevance in socioeconomic analyses.
Additionally, our research results could be used for a comparative analysis of the performance of different Gradient Boosting hyperparameter tuning approaches.
Furthermore, the integration of parameter tuning within the Expectation-Maximization (EM) algorithm, as well as an integration in the bootstrap in the MSE estimations should be discussed. However, it is essential to note that such integration may introduce computational challenges due to increased processing time, which requires careful consideration. For a better comparison, the approach should also be compared to tuned Random Forest (RF) models.

Another possible research aspect of interest could be to include further machine learning methods in the EM algorithm, such as Support Vector Machines (SVMs) \citep{Krennmair2022}. Because of SVMs capability to handle nonlinear data through the use of kernel methods, their inclusion warrants further examination in future studies.

\bibliographystyle{chicago}
\bibliography{Datenbank}

\begin{thebibliography}{}

\bibitem[\protect\citeauthoryear{Akaike}{Akaike}{1973}]{Akaike1973}
Akaike, H. (1973).
\newblock Information theory and an extension of the maximum likelihood
  principle.
\newblock In F.~Csaki and B.~N. Petrov (Eds.), {\em Information Theory:
  Proceedings of the 2nd International Symposium}, pp.\  267--281. Akademiai
  Kiado, Budapest.

\bibitem[\protect\citeauthoryear{Akaike}{Akaike}{1978}]{Akaike1978}
Akaike, H. (1978).
\newblock On the likelihood of a time series model.
\newblock {\em Journal of the Royal Statistical Society. Series D (The
  Statistician)\/}~{\em 27\/}(3/4), 217--235.

\bibitem[\protect\citeauthoryear{Cavanaugh and Neath}{Cavanaugh and
  Neath}{2019}]{Cavanaugh2019}
Cavanaugh, J.~E. and A.~A. Neath (2019).
\newblock The akaike information criterion: Background, derivation, properties,
  application, interpretation, and refinements.
\newblock {\em WIREs Computational Statistics\/}~{\em 11\/}(3), 1--11.

\bibitem[\protect\citeauthoryear{Liang, Wu, and Zou}{Liang
  et~al.}{2008}]{Liang2008}
Liang, H., H.~Wu, and G.~Zou (2008).
\newblock A note on conditional aic for linear mixed-effects models.
\newblock {\em Biometrika\/}~{\em 95\/}(3), 773--778.

\bibitem[\protect\citeauthoryear{Vaida and Blanchard}{Vaida and
  Blanchard}{2005}]{Vaida2005}
Vaida, F. and S.~Blanchard (2005).
\newblock Conditional akaike information for mixed-effects models.
\newblock {\em Biometrika\/}~{\em 92\/}(2), 351--370.

\end{thebibliography}


\begin{thebibliography}{}

\bibitem[\protect\citeauthoryear{Battese, Harter, and Fuller}{Battese
  et~al.}{1988}]{battese}
Battese, G., R.~Harter, and W.~Fuller (1988).
\newblock An error-components model for prediction of county crop areas using
  survey and satellite data.
\newblock {\em Journal of the American Statistical Association\/}~{\em
  83\/}(401), 28--36.

\bibitem[\protect\citeauthoryear{Breiman}{Breiman}{1999}]{breiman_predictiongames}
Breiman, L. (1999).
\newblock Prediction games and arcing algorithms.
\newblock {\em Neural Computation\/}~{\em 11\/}(7), 1493--1517.

\bibitem[\protect\citeauthoryear{Carlstein, Do, Hall, Hesterberg, and
  Künsch}{Carlstein et~al.}{1998}]{carlstein1998matched}
Carlstein, E., K.-A. Do, P.~Hall, T.~Hesterberg, and H.~R. Künsch (1998).
\newblock Matched-block bootstrap for dependent data.
\newblock {\em Bernoulli\/}~{\em 4\/}(3), 305--328.

\bibitem[\protect\citeauthoryear{Chambers and Chandra}{Chambers and
  Chandra}{2013}]{chambers_bootstrap}
Chambers, R. and H.~Chandra (2013).
\newblock A random effect block bootstrap for clustered data.
\newblock {\em Journal of Computational and Graphical Statistics\/}~{\em
  22\/}(2), 452--470.

\bibitem[\protect\citeauthoryear{Chen and Guestrin}{Chen and
  Guestrin}{2016}]{xgboost_paper}
Chen, T. and C.~Guestrin (2016).
\newblock Xgboost: A scalable tree boosting system.
\newblock In {\em Proceedings of the 22nd ACM SIGKDD International Conference
  on Knowledge Discovery and Data Mining}, KDD '16, New York, NY, USA, pp.\
  785--794. Association for Computing Machinery.

\bibitem[\protect\citeauthoryear{Chen, He, Benesty, Khotilovich, Tang, Cho,
  Chen, Mitchell, Cano, Zhou, Li, Xie, Lin, Geng, Li, and Yuan}{Chen
  et~al.}{2022}]{xgboost_paket}
Chen, T., T.~He, M.~Benesty, V.~Khotilovich, Y.~Tang, H.~Cho, K.~Chen,
  R.~Mitchell, I.~Cano, T.~Zhou, M.~Li, J.~Xie, M.~Lin, Y.~Geng, Y.~Li, and
  J.~Yuan (2022).
\newblock {\em xgboost: Extreme Gradient Boosting}.
\newblock R package version 1.6.0.1.

\bibitem[\protect\citeauthoryear{Dempster, Laird, and Rubin}{Dempster
  et~al.}{1977}]{em_autoren}
Dempster, A.~P., N.~M. Laird, and D.~B. Rubin (1977).
\newblock Maximum likelihood from incomplete data via the em algorithm.
\newblock {\em Journal of the Royal Statistical Society: Series B
  (Methodological)\/}~{\em 39\/}(1), 1--22.

\bibitem[\protect\citeauthoryear{Fahrmeir, Kneib, Lang, and Marx}{Fahrmeir
  et~al.}{2013}]{regression_fahrmeir}
Fahrmeir, L., T.~Kneib, S.~Lang, and B.~Marx (2013).
\newblock {\em Regression: Models, Methods and Applications}.
\newblock Berlin: Springer-Verlag.

\bibitem[\protect\citeauthoryear{Freund and Schapire}{Freund and
  Schapire}{1997}]{boosting_idea}
Freund, Y. and R.~E. Schapire (1997).
\newblock A decision-theoretic generalization of on-line learning and an
  application to boosting.
\newblock {\em Journal of Computer and System Sciences\/}~{\em 55\/}(1),
  119--139.

\bibitem[\protect\citeauthoryear{Friedman}{Friedman}{2001}]{friedman_gb}
Friedman, J.~H. (2001).
\newblock Greedy function approximation: A gradient boosting machine.
\newblock {\em The Annals of Statistics\/}~{\em 29\/}(5), 1189--1232.

\bibitem[\protect\citeauthoryear{Hajjem, Bellavance, and Larocque}{Hajjem
  et~al.}{2014}]{MERF}
Hajjem, A., F.~Bellavance, and D.~Larocque (2014).
\newblock Mixed-effects random forest for clustered data.
\newblock {\em Journal of Statistical Computation and Simulation\/}~{\em
  84\/}(6), 1313--1328.

\bibitem[\protect\citeauthoryear{Horvitz and Thompson}{Horvitz and
  Thompson}{1952}]{horvitz1952generalization}
Horvitz, D.~G. and D.~J. Thompson (1952).
\newblock A generalization of sampling without replacement from a finite
  universe.
\newblock {\em Journal of the American Statistical Association\/}~{\em
  47\/}(260), 663--685.

\bibitem[\protect\citeauthoryear{Krennmair}{Krennmair}{2022}]{SAEforest}
Krennmair, P. (2022).
\newblock {\em SAEforest: Mixed Effect Random Forests for Small Area
  Estimation}.
\newblock R package version 0.1.0.

\bibitem[\protect\citeauthoryear{Krennmair and Schmid}{Krennmair and
  Schmid}{2022}]{Krennmair2022}
Krennmair, P. and T.~Schmid (2022).
\newblock Flexible domain prediction using mixed effects random forests.
\newblock {\em Journal of the Royal Statistical Society: Series C (Applied
  Statistics)\/}~{\em 71\/}(5), 1865--1894.

\bibitem[\protect\citeauthoryear{Kreutzmann, Pannier, Rojas-Perilla, Schmid,
  Templ, and Tzavidis}{Kreutzmann et~al.}{2019}]{emdi}
Kreutzmann, A.-K., S.~Pannier, N.~Rojas-Perilla, T.~Schmid, M.~Templ, and
  N.~Tzavidis (2019).
\newblock The {R} package {emdi} for estimating and mapping regionally
  disaggregated indicators.
\newblock {\em Journal of Statistical Software\/}~{\em 91\/}(7), 1--33.

\bibitem[\protect\citeauthoryear{Merfeld and Newhouse}{Merfeld and
  Newhouse}{2023}]{merfeld_newhouse_xgboost}
Merfeld, J.~D. and D.~Newhouse (2023).
\newblock Improving estimates of mean welfare and uncertainty in developing
  countries.
\newblock World Bank Policy Research Working Paper 10348, World Bank.

\bibitem[\protect\citeauthoryear{Molina and Marhuenda}{Molina and
  Marhuenda}{2015}]{sae}
Molina, I. and Y.~Marhuenda (2015).
\newblock {sae}: An {R} package for small area estimation.
\newblock {\em The R Journal\/}~{\em 7\/}(1), 81--98.

\bibitem[\protect\citeauthoryear{Molina and Rao}{Molina and
  Rao}{2010}]{molina2009small}
Molina, I. and J.~N. Rao (2010).
\newblock Small area estimation of poverty indicators.
\newblock {\em Canadian Journal of statistics\/}~{\em 38\/}(3), 369--385.

\bibitem[\protect\citeauthoryear{Morales, Esteban, Pérez, and Hobza}{Morales
  et~al.}{2021}]{Morales2021SmallAreaEstimation}
Morales, D., M.~D. Esteban, A.~Pérez, and T.~Hobza (2021).
\newblock {\em A Course on Small Area Estimation and Mixed Models: Methods,
  Theory and Applications in R}.
\newblock Statistics for Social and Behavioral Sciences. Cham: Springer.

\bibitem[\protect\citeauthoryear{{R Core Team}}{{R Core Team}}{2023}]{R}
{R Core Team} (2023).
\newblock {\em R: A Language and Environment for Statistical Computing}.
\newblock Vienna, Austria: R Foundation for Statistical Computing.

\bibitem[\protect\citeauthoryear{Rojas-Perilla, Pannier, Schmid, and
  Tzavidis}{Rojas-Perilla et~al.}{2020}]{rojas2020data}
Rojas-Perilla, N., S.~Pannier, T.~Schmid, and N.~Tzavidis (2020).
\newblock Data-driven transformations in small area estimation.
\newblock {\em Journal of the Royal Statistical Society: Series A (Statistics
  in Society)\/}~{\em 183\/}(1), 121--148.

\end{thebibliography}

\newpage
\section*{Appendix}
\label{Anhang}
The figures displayed in the Appendix include the tuned MEGB (MEGB Tuned), which was tuned using a systematic sequential grid search parameter tuning. This has the important advantage of enabling the reduction of dimensions. If all parameters were tuned at the same time and all combinations were tested at the same time, the dimension of the grid would be enormous. The presented approach can therefore be seen as a reduced grid search.

\begin{figure}[htb!]
    \centering
    \includegraphics[width = \linewidth]{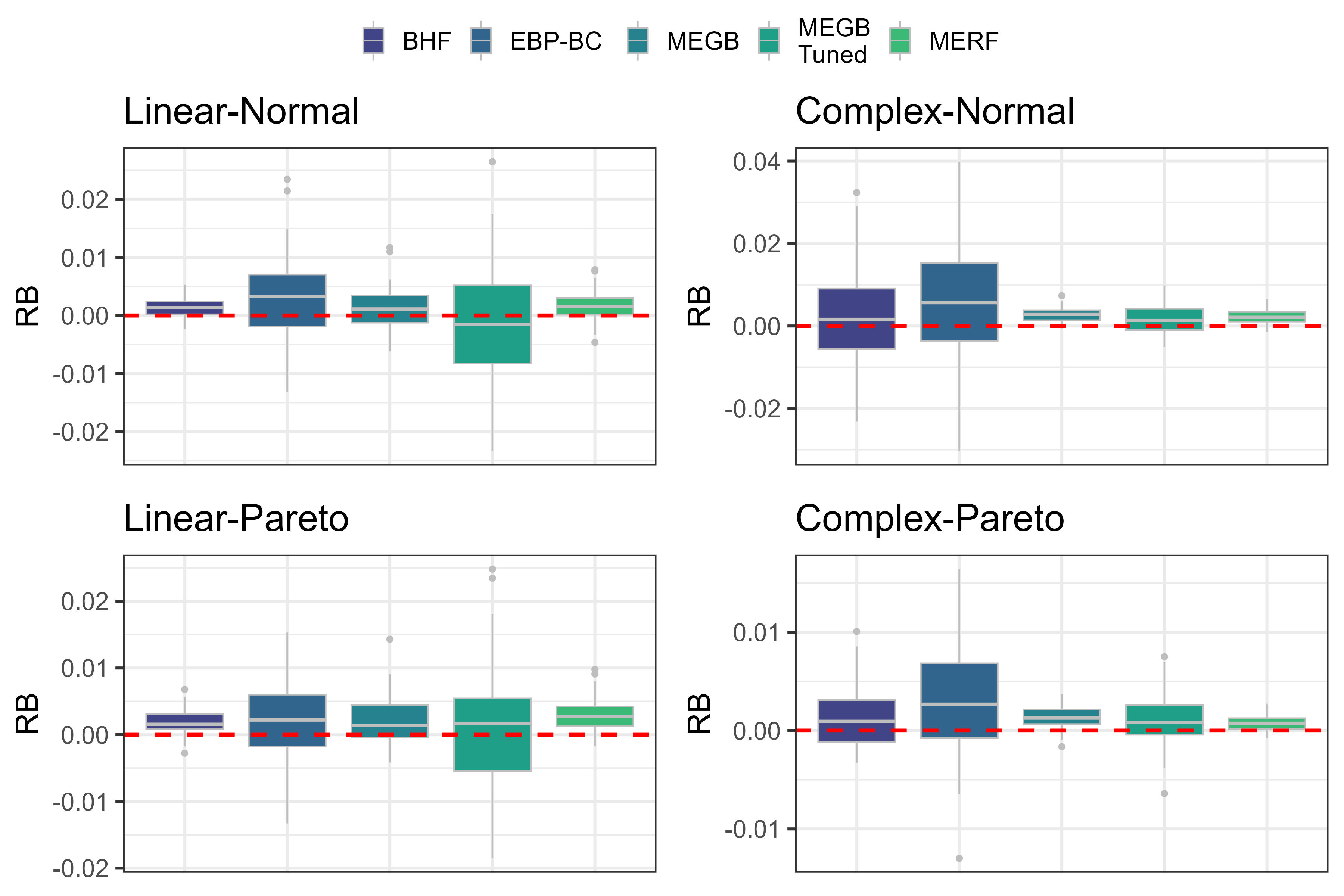}
    \caption{Relative bias averaged over 20 MC simulation runs in the model-based simulation study for the four
scenarios}
    \label{fig:mb bias tuned}
\end{figure}

\begin{table} [htb!]
\centering
\begin{tabular}{lllllllll}
\toprule
\multicolumn{1}{c}{RB (\%)} & \multicolumn{2}{c}{Linear-Normal} & \multicolumn{2}{c}{Complex-Normal} & \multicolumn{2}{c}{Linear-Pareto} & \multicolumn{2}{c}{Complex-Pareto} \\
\cmidrule(l{3pt}r{3pt}){1-1} \cmidrule(l{3pt}r{3pt}){2-3} \cmidrule(l{3pt}r{3pt}){4-5} \cmidrule(l{3pt}r{3pt}){6-7} \cmidrule(l{3pt}r{3pt}){8-9}
  & Mean & Median & Mean & Median & Mean & Median & Mean & Median\\
\midrule
BHF & 0.133 & 0.133 & 0.239 & 0.161 & 0.190 & 0.157 & 0.127 & 0.094\\
EBP-BC & 0.279 & 0.329 & 0.576 & 0.565 & 0.190 & 0.221 & 0.308 & 0.268\\
MEGB & 0.127 & 0.113 & 0.260 & 0.280 & 0.207 & 0.140 & 0.134 & 0.127\\
MEGB
Tuned & -0.068 & -0.151 & 0.173 & 0.134 & 0.063 & 0.169 & 0.109 & 0.082\\
MERF & 0.166 & 0.156 & 0.237 & 0.210 & 0.310 & 0.277 & 0.072 & 0.074\\
\bottomrule
\end{tabular}
\label{mb_rb_tuned_tab}
\caption{Mean and the median  of the relative bias in \% for the estimators across the four scenarios in the model-based simulation study}
\end{table}

\begin{figure}[htb!]
    \centering
    \includegraphics[width = \linewidth]{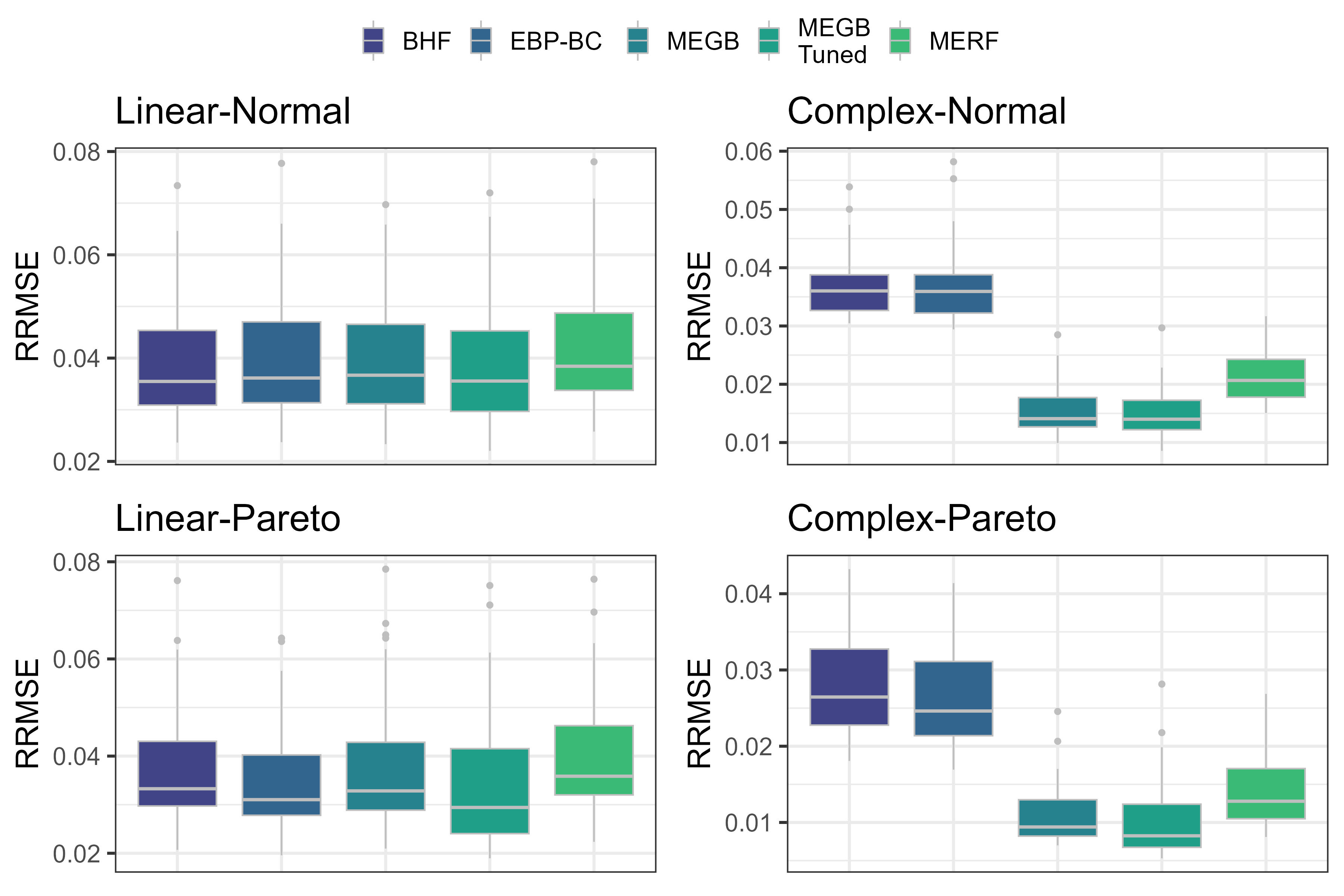}
    \caption{RRMSE averaged over 20 MC simulation runs in the model-based simulation study for the four scenarios}
    \label{fig:mb rmse tuned}
\end{figure}

\begin{table}[htb!]
\centering
\begin{tabular}{lllllllll}
\toprule
\multicolumn{1}{c}{RRMSE (\%)} & \multicolumn{2}{c}{Linear-Normal} & \multicolumn{2}{c}{Complex-Normal} & \multicolumn{2}{c}{Linear-Pareto} & \multicolumn{2}{c}{Complex-Pareto} \\
\cmidrule(l{3pt}r{3pt}){1-1} \cmidrule(l{3pt}r{3pt}){2-3} \cmidrule(l{3pt}r{3pt}){4-5} \cmidrule(l{3pt}r{3pt}){6-7} \cmidrule(l{3pt}r{3pt}){8-9}
  & Mean & Median & Mean & Median & Mean & Median & Mean & Median\\
\midrule
BHF & 3.907 & 3.548 & 3.652 & 3.602 & 3.770 & 3.328 & 2.755 & 2.645\\
EBP-BC & 3.986 & 3.614 & 3.660 & 3.593 & 3.512 & 3.103 & 2.630 & 2.462\\
MEGB & 3.982 & 3.669 & 1.544 & 1.410 & 3.768 & 3.284 & 1.088 & 0.941\\
MEGB
Tuned & 3.891 & 3.556 & 1.504 & 1.401 & 3.445 & 2.943 & 1.011 & 0.825\\
MERF & 4.187 & 3.842 & 2.171 & 2.066 & 4.021 & 3.586 & 1.372 & 1.278\\
\bottomrule
\end{tabular}
\label{mb_rmse_tuned_tab}
\caption{Mean and the median Relative RMSE in \% for the estimators across the four scenarios in the model-based simulation study}
\end{table}

\begin{figure}[htb!]
    \centering
    \includegraphics[width = \linewidth]{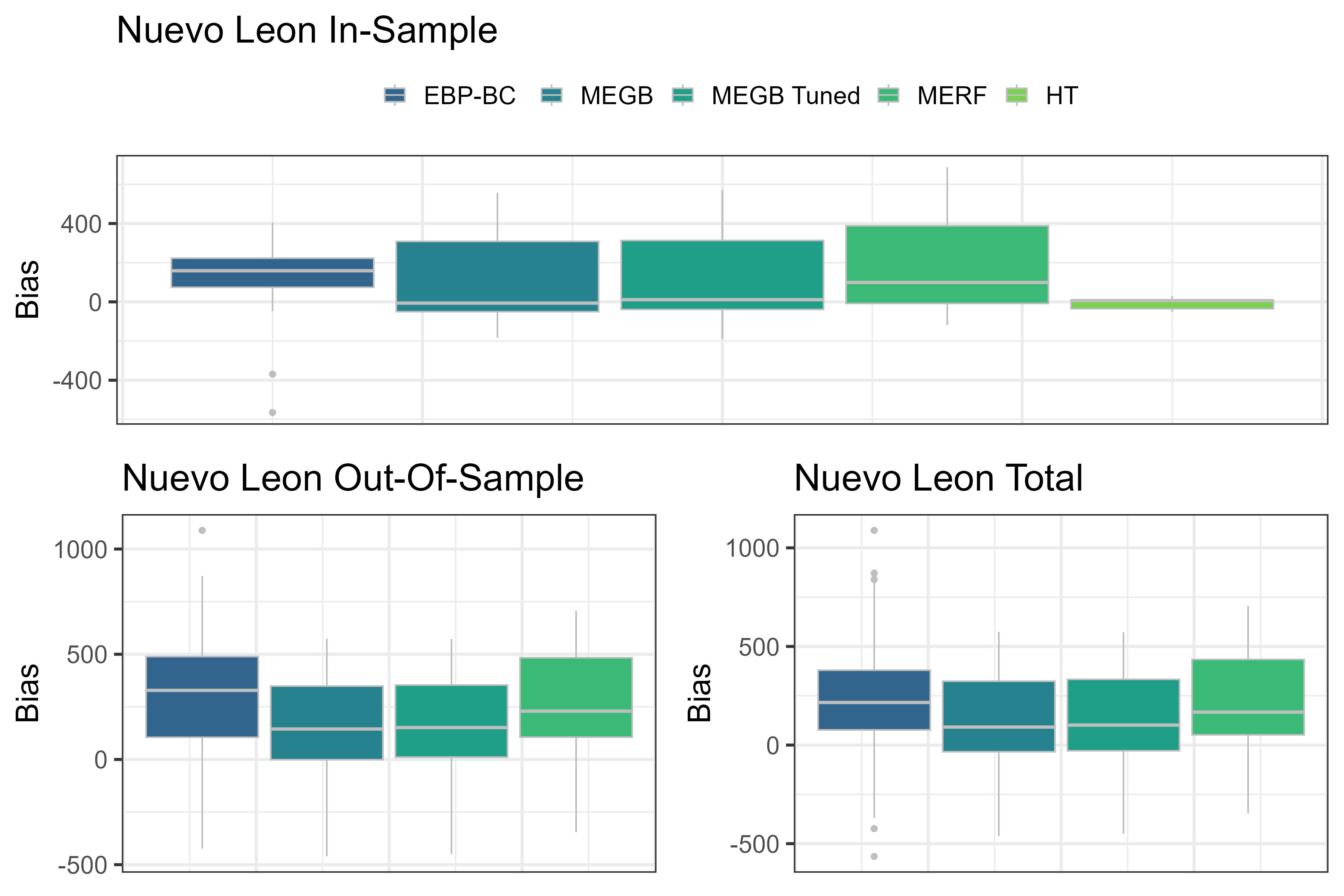}
    \caption{Bias for In-Sample -, Out-Of-Sample- and Total municipalities in Nuevo Léon for the EBP-BC, MEGB, MEGB Tuned, MERF and the direct estimator \citep{horvitz1952generalization}}
    \label{db bias tuned}
\end{figure}

\begin{table}[htb!]
\centering
\begin{tabular}{lllllll}
\toprule
\multicolumn{1}{c}{Bias} & \multicolumn{2}{c}{In-Sample} & \multicolumn{2}{c}{Out-Of-Sample} & \multicolumn{2}{c}{Total} \\
\cmidrule(l{3pt}r{3pt}){1-1} \cmidrule(l{3pt}r{3pt}){2-3} \cmidrule(l{3pt}r{3pt}){4-5} \cmidrule(l{3pt}r{3pt}){6-7}
  & Mean & Median & Mean & Median & Mean & Median\\
\midrule
EBP-BC & 109.286 & 158.782 & 319.249 & 328.412 & 232.794 & 215.931\\
MEGB & 102.209 & -6.291 & 157.357 & 144.780 & 134.649 & 91.270\\
MEGB
Tuned & 102.547 & 11.135 & 157.563 & 151.590 & 134.909 & 101.158\\
MERF & 188.521 & 99.188 & 255.364 & 229.452 & 227.841 & 167.026\\
\bottomrule
\end{tabular}
\label{tab_db_bias_tuned}
\caption{Mean and median bias for In-Sample-, Out-Of-Sample-, and Total municipalities
in Nuevo Léon for the EBP-BC, MEGB, MEGB Tuned and the MERF of the design-based simulation study}
\end{table}

\begin{table}[htb!]
\centering
\begin{tabular}{lllllll}
\toprule
\multicolumn{1}{c}{RMSE} & \multicolumn{2}{c}{In-Sample} & \multicolumn{2}{c}{Out-Of-Sample} & \multicolumn{2}{c}{Total} \\
\cmidrule(l{3pt}r{3pt}){1-1} \cmidrule(l{3pt}r{3pt}){2-3} \cmidrule(l{3pt}r{3pt}){4-5} \cmidrule(l{3pt}r{3pt}){6-7}
  & Mean & Median & Mean & Median & Mean & Median\\
\midrule
EBP-BC & 316.664 & 302.027 & 417.592 & 346.681 & 376.034 & 327.724\\
MEGB & 219.948 & 138.814 & 257.417 & 207.796 & 241.989 & 189.206\\
MEGB
Tuned & 217.871 & 143.017 & 257.700 & 215.551 & 241.300 & 195.441\\
MERF & 256.574 & 127.962 & 316.186 & 272.137 & 291.640 & 208.215\\
\bottomrule
\end{tabular}
\label{tab_db_rmse_tuned}
\caption{Mean and median RMSE for In-Sample-, Out-Of-Sample-, and Total municipalities
in Nuevo Léon for the EBP-BC, MEGB, MEGB Tuned and the MERF of the design-based simulation study}
\end{table}

\begin{figure}[htb!]
    \centering
    \includegraphics[width = \linewidth]{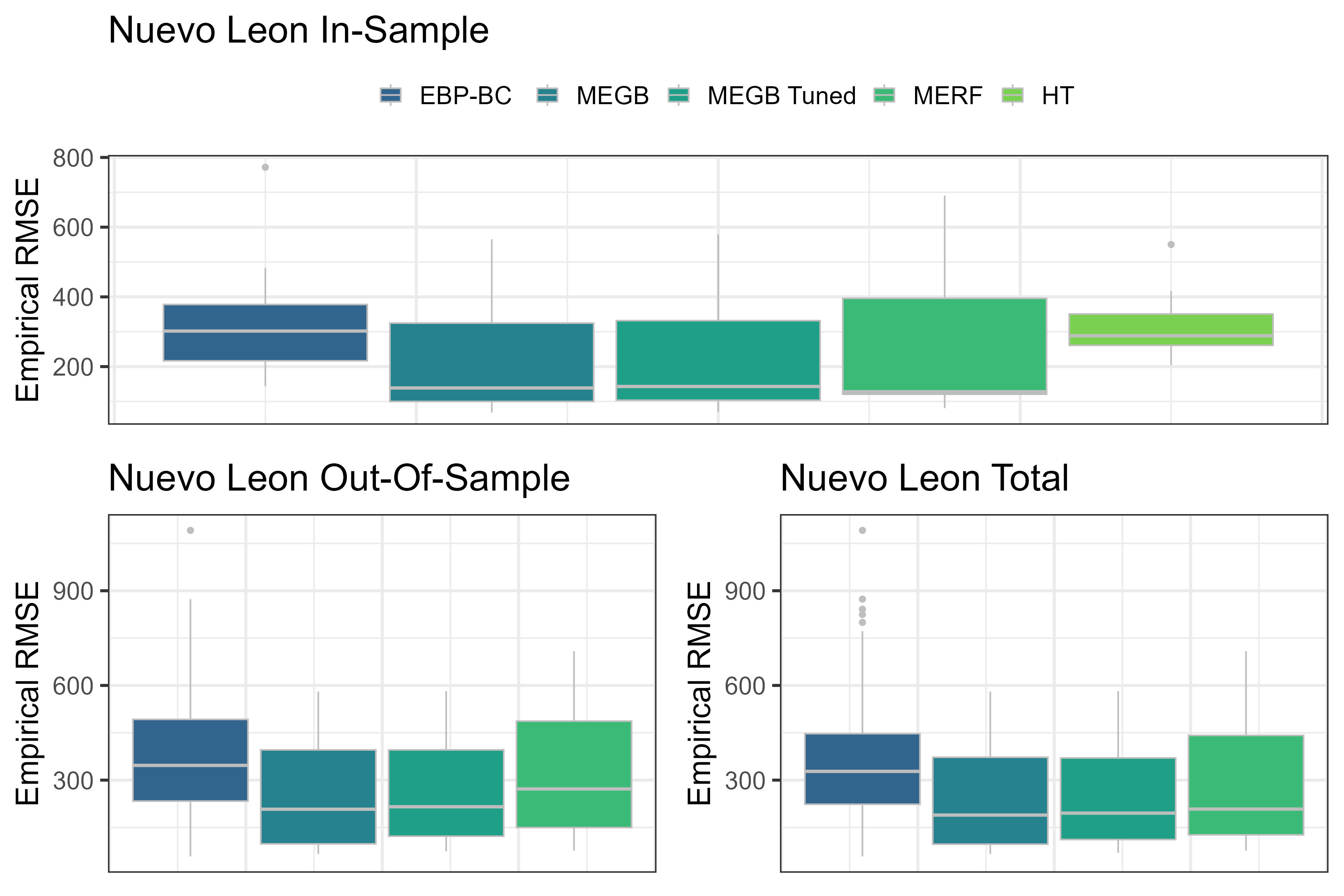}
    \caption{RMSE averaged over 50 MC simulation runs for In-Sample -, Out-Of-Sample- and Total municipalities in Nuevo Léon for the EBP-BC, MEGB, MEGB Tuned, MERF and the Horvitz-Thompson estimator}
    \label{fig:rrmse_db}
\end{figure}

\end{document}